\documentclass[aps,prd,twocolumn,groupedaddress,amssymb,eqsecnum,showpacs,epsfig,nofootinbib]{revtex4-1}
\usepackage{graphicx}
\usepackage{bm}
\usepackage{dcolumn}
\usepackage{amsmath}
\usepackage{verbatim}
\usepackage{amssymb}

\numberwithin{equation}{section}
\usepackage{epsfig}

\def \lleq {\lower0.9ex\hbox{ $\buildrel < \over \sim$} ~}
\def \ggeq {\lower0.9ex\hbox{ $\buildrel > \over \sim$} ~}

\def \beq  {\begin{equation}}
\def \eeq  {\end{equation}}
\def \ber  {\begin{eqnarray}}
\def \eer  {\end{eqnarray}}

\usepackage[usenames]{color}

\newcommand{\newc}{\newcommand}

\newc{\diag}{\mathop{\mathrm{diag}}}
\newc{\be}{\begin{equation}}
\newc{\ee}{\end{equation}}
\newc{\ba}{\begin{eqnarray}}
\newc{\ea}{\end{eqnarray}}
\newc{\bea}{\begin{eqnarray*}}
\newc{\eea}{\end{eqnarray*}}
\newc{\D}{\partial}
\newc{\ie}{{\it i.e.} }
\newc{\eg}{{\it e.g.} }
\newc{\etc}{{\it etc.} }
\newc{\etal}{{\it et al.}}
\newc{\lcdm }{$\Lambda$CDM }

\newc{\ra}{\rightarrow}
\newc{\lra}{\leftrightarrow}
\newc{\lsim}{\buildrel{<}\over{\sim}}
\newc{\gsim}{\buildrel{>}\over{\sim}}

\begin{document}

\title{Generalized LTB model with Inhomogeneous Isotropic Dark Energy: Observational Constraints}
\author{J.~Grande$^a$ and L.~Perivolaropoulos$^b$}
 \affiliation{$^a$ Dept.~d'ECM and Institut de Ci\`encies del Cosmos,
Univ.~de Barcelona, Av.~Diagonal 647, E-08028 Barcelona, Spain\\$^b$Department of Physics, University of Ioannina, Greece}
\date{\today}

\begin{abstract}
We consider on-center and off-center observers in an inhomogeneous, spherically symmetric, isocurvature (flat) concentration of dark energy with typical size of a few Gpc. Such a concentration could be produced e.g. by a recently formed global monopole with core size that approaches the Hubble scale. In this case we would have what may be called `topological quintessence' in analogy with the well-known \emph{topological inflation}. We show that the minimum comoving radius $r_{0min}$ of such a dark energy inhomogeneity that is consistent with the Union2 Type Ia supernovae (SnIa) data at the $3\sigma$ level is $r_{0min}\simeq 1.8{\rm \,Gpc}$. As expected, the best-fit fractional dark energy density at the center, $\Omega_{X, \text{in}}$, approaches the corresponding \lcdm value $\Omega_{X, \text{in}} =0.73$ for large enough values of the inhomogeneity radius $r_0$ ($r_0>4{\rm \,Gpc}$).
Using the Union2 data, we show that the maximum allowed shift $r_{obs-max}$ of the observer from the center of the inhomogeneity is about $0.7 r_0$ which respects the Copernican principle. The model naturally predicts the existence of a preferred axis and alignment of the low CMB multipoles. However, the constraints on $r_{obs-max}$ coming from the magnitude of the CMB dipole remain a severe challenge to the Copernican principle and lead to $r_{obs-max}< 110{\rm \,Mpc}$ even for an inhomogeneity radius as large as $r_0=7{\rm \,Gpc}$.
\end{abstract}
\maketitle

\section{Introduction}
Detailed cosmological observations made during the last decade (see~\cite{Komatsu:2010fb} for some of the latest results) have made it clear that the minimal cosmological model (sCDM) based on the cosmological principle (homogeneity, isotropy, validity of General Relativity) and with a purely radiation and matter content (baryonic and Cold Dark Matter (CDM)), is inconsistent with observations. The simplest extension of sCDM able to reconcile it with the observed accelerated cosmic expansion (\lcdm) involves the existence of a homogeneous, constant in time energy density that constitutes about $70\%$ of the total energy density of the universe and is known as the {\it cosmological constant}~\cite{lcdmrev}. A time-dependent generalization of this energy has been dubbed {\it dark energy}. The repulsive gravitational properties of the cosmological constant dominate on large %cosmological
scales and may induce the observed acceleration. Due to its simplicity and its consistency with most cosmological observations \lcdm is currently the {\it standard} cosmological model, replacing sCDM, which held that status in the early 90's, when observational data were far less extensive and accurate.

However,  \lcdm is also faced with some challenges, which originate from both theoretical and observational considerations. The two main theoretical challenges of the model are {\it fine tuning} and {\it lack of theoretical motivation}. Indeed, in order to make the model consistent with observations, the energy scale of the cosmological constant must be fine-tuned to an unnaturally small value, $120$ orders of magnitude smaller than the one anticipated from theoretical models of quantum field theory~\cite{lcdmrev}.

From the observational point of view, even though \lcdm is %consistent
in agreement with the vast majority of data, there is a set of observations that appear to be mildly inconsistent with the model at the $2-3 \sigma$ level. Most of these observations, which we proceed to summarize, seem to be related to the existence of a preferred cosmological axis:
\begin{itemize}
\item
{\bf Planarity and Alignment of the CMB Low Multipole Moments:} It has been known since 2003~\cite{Tegmark:2003ve,Land:2005ad} (see~\cite{Copi:2010na,Bennett:2010jb} for recent reviews) that the quadrupole and octopole components of the CMB temperature perturbation maps (the largest scale part of those maps) are dominated by planar features which are unnaturally aligned with each other in the direction (perpendicular to the planes) $(l,b)\simeq (250^\circ,60^\circ)$~\cite{Park:2006dv,Abramo:2006gw,Gruppuso:2010up}. The probability that such combination of planarity and alignment occurs in a random Gaussian map is less than $0.5\%$~\cite{Copi:2010na,Bennett:2010jb}. This peculiar feature has been attributed to a posteriori considerations~\cite{Bennett:2010jb}, namely {\it within a large number of features of a map, it is statistically anticipated that some of them will present large statistical fluctuations}.  This type of argument would be stronger if the large fluctuations affected random uncorrelated features (for example multipoles with $l=53$ and $l=79$). In that case we would have an a posteriori arbitrary selection of features. The quadrupole-octopole moments, however, constitute \emph{all} the information we have about the CMB temperature fluctuation maps on the largest angular scales. Thus the observed alignment could also be restated by saying that {\it the large scale features of the CMB maps are significantly more planar than anticipated for random Gaussian maps}. The fact that the {\it preferred} plane is relatively close to the galactic plane (which is usually masked due to noise, before the construction of the power spectrum is made) is probably responsible~\cite{Pontzen:2010yw} for the relatively low values of the angular power spectrum components $C_2$, $C_3$, which is considered by some authors to be an independent puzzle~\cite{Rakic:2007ve}.
\item
{\bf Large Scale Velocity Bulk Flows:} Recent studies~\cite{Watkins:2008hf,Kashlinsky:2008ut,Lavaux:2008th} have detected a large scale flow with a bulk velocity
in the range $[400, 1000]$~km/s towards $(l,b)\simeq (282^\circ,8^\circ)$ on scales $100-300 {\rm \,Mpc}$. This detection is larger than the \lcdm prediction and inconsistent with it at the level of $2-3\sigma$.
\item
{\bf Large scale alignment in the QSO optical polarization data:} Quasar polarization vectors are not randomly oriented over the sky with a probability often in excess of 99.9\%. The alignment effect seems to be prominent along a particular axis with direction $(l,b)=(267^\circ, 69^\circ)$~\cite{Hutsemekers:2005iz}.
\item
{\bf Profiles of Cluster Haloes:} \lcdm predicts shallow, low concentration and density profiles, in contrast to some observations which point to denser 
cluster haloes~\cite{Broadhurst:2004bi,Umetsu:2007pq}.
\end{itemize}

As announced, three of the above four puzzles are large-scale effects related to preferred cosmological directions (CMB multipole alignments, QSO polarization alignment and large scale bulk flows); those directions, moreover, appear to be not far from each other, being approximately normal to the axis of the ecliptic poles $(l,b)=(96^\circ,30^\circ)$ and lying close to the ecliptic plane and the equinoxes. This coincidence has triggered investigations for possible systematic effects related to the CMB preferred axis but no significant such effects have been found~\cite{Copi:2010na}.

Finally, it should be mentioned that, in addition to the discussed large-scale effects, the \lcdm model also faces some issues on galactic scales (missing satellites problem~\cite{Klypin:1999uc,Moore:2001fc,Madau:2008fr} and the cusp/core nature
of the central density profiles of dwarf galaxies~\cite{cuspygal}).

In spite of those problems, the construction of generalizations of \lcdm has been motivated so far mainly by theoretical rather than observational considerations. Some of the most popular ideas among those that go beyond \lcdm are the following:

\begin{itemize}
\item
{\bf Time Dependent Dark Energy~\cite{Copeland:2006wr}:} Within this approach, dark energy retains its homogeneous and isotropic character, but may present a time dependence. This time dependence aims at solving (or at least alleviating) the previously mentioned fine-tuning problem, i.e. the unnaturally small value of the current dark energy required by the observations. Indeed, it turns out that a dark energy density which is larger at early times and evolves via proper dynamics to its present low value is easier to accommodate in theoretical models. A representative of this class of models is {\it quintessence}, where the role of dark energy is played by a dynamically evolving scalar field. However, for this evolution to be consistent with the observations, the scalar field must be assigned an extremely (fine-tuned) small mass, fact that
makes the additional complexity introduced in this class of models be of questionable value.
\item
{\bf Modified Gravity:} The repulsive gravity provided by the cosmological constant on large scales can also be induced by modifications of General Relativity in the context of a homogeneous and isotropic background. Several such modifications have been suggested, including scalar tensor gravity~\cite{stgrav}, DGP models that involve gravitons leaking through large extra dimensions~\cite{Lue:2005ya}, $f(R)$ gravity~\cite{frgrav} etc. These models are well motivated as complete geometric physical theories, but they have more degrees of freedom than \lcdm and they also require significant fine-tuning in order to be consistent with both the observed cosmic accelerating expansion and the gravitational solar system constraints that prevent large local departures from General Relativity.
\item
{\bf Inhomogeneous and Isotropic Lemaitre-Tolman-Bondi Void Models~\cite{Lemaitre:1933qe,Tolman:1934za,Bondi:1947av}:} Within this framework, an additional dark energy component is not needed to secure consistency with the cosmological data that indicate accelerating expansion. The basic idea that lies behind them is to consider that the increased expansion rate occurs locally in space rather than at recent cosmological times, a fact that can be achieved by assuming a locally-reduced matter energy density~\cite{Krasinski}. Thus, the observer is placed close to the center of a giant void with dimensions of a few Gpc~\cite{Alnes:2005rw}. Even though this approach is free of dark energy, it is by no means free of fine tuning. Apart from the unnatural assumption of giant-size {\rm \,Gpc} voids, which are very unlikely to be produced in any cosmology, these models require the observer to be placed within a very small volume at the center of the void (about $10^{-6}$ of the total volume of the void). A slightly off-center observer, however, will naturally experience a preferred cosmological direction (towards the center of the void), which may help to resolve some of the observational puzzles of \lcdm discussed above.
\end{itemize}
In the present study we consider yet an alternative approach for the generalization of \lcdm. Instead of breaking the time translation invariance of dark energy, as in the class of quintessence models, we break its space translation invariance (homogeneity). We thus consider a spherically symmetric overdensity of dark energy with typical scale of a few {\rm \,Gpc}: {\it Inhomogeneous Dark Energy (IDE)}. This class of models adds degrees of freedom to \lcdm in a manner that is generically different from the other three classes considered above. Thus, it also has specific advantages as compared to them. In particular:
\begin{itemize}
\item
{\bf Physical Mechanism:} There is a well-defined {\it physical mechanism} that can produce this type of spherically symmetric, Hubble-scale IDE. It is based on applying the principles of {\it topological inflation}~\cite{Vilenkin:1994pv} to the case of late-time acceleration. According to the idea of topological inflation, the false vacuum energy of the core of a topological defect can give rise to accelerating expansion if the core size reaches the Hubble scale when gravity starts dominating the dynamics. Thus, for example, a recently formed global monopole with appropriate scale of symmetry breaking and coupling could naturally produce a Hubble-scale, spherically symmetric, isocurvature dark energy overdensity. By analogy with topological inflation, this mechanism may be called {\it topological quintessence}.
\item
{\bf Naturally Large Scale:} Whereas matter inhomogeneities (voids) of a few Gpc are very unlikely to be formed through large scale structure mechanisms, IDE on Hubble scales is more natural and easy to construct. In the IDE case (also in contrast to inhomogeneous matter models), the homogeneous limit (\lcdm) is consistent with observations and one interesting question is {\it `What is the smallest scale of inhomogeneity for which this consistency remains?'}.
\item
{\bf More Natural Off-center Observers (less fine tuning):} Due to the arbitrarily large scale of IDE (supported by observations, as we will see), there is less fine tuning for the location of the observer with respect to the center of the spherical inhomogeneity. In fact, as the inhomogeneity scale approaches the horizon the model becomes indistinguishable from \lcdm and there is no observational constraint for the location of the observer. %with respect to the center.
\item
{\bf Natural Generation of a Preferred Cosmological Axis:} As we have seen, most of the observational challenges of \lcdm hint towards the existence of a preferred cosmological direction. Such a direction is naturally provided for an off-center observer in an inhomogeneous matter or dark energy model: {\it the direction that connects the location of the observer with the center of the inhomogeneity}. The displacement of the observer naturally leads to an alignment of low CMB multipole moments and bulk velocity flows.
\end{itemize}
The additional degrees of freedom (with respect to \lcdm) introduced in this class of models are similar to the corresponding ones in models that break the time translation invariance of dark energy. We also need a scalar field, but now with spatial rather than temporal dependence. Compared to matter void models, however, the IDE setup requires more (and less natural) degrees of freedom. This is the price to pay for the reduction in the amount of fine tuning.

The goal of the present study is to investigate some of the observational consequences and constraints of IDE models. We focus on the constraints coming from SnIa data and the CMB low multipole moments. We also discuss the physical mechanism that could give rise to IDE (\emph{topological quintessence}), although we postpone a complete analysis of this mechanism for a future study~\cite{inprog}.

The structure of this paper is the following: In the next section we derive the cosmological equations, light-like geodesics and luminosity distance for on and off-center observers in the presence of IDE. We also discuss the basics of topological quintessence as a mechanism for producing inhomogeneous dark energy and derive the angular dependence of the CMB temperature fluctuations induced by the shift of the observer from the center of the inhomogeneity. In section 3 we use the predicted luminosity distance to constrain the model parameters using the Union2 SnIa data~\cite{Amanullah:2010vv}. In the same section we compare the values of the low multipole moments (dipole, quadrupole) of CMB perturbations induced by the displacement of the observer from the center to the corresponding observed values and derive constraints on the parameters of the model. Finally, in section 4 we conclude and discuss the implications and possible extensions of our results. In what follows we fix the present age of the universe to $t_0=13.7$\,Gyr. In the best fit \lcdm model, this would correspond to fixing $H_0=71$\,km/(s\,Mpc). We also refer to a comoving radial distance simply as `distance'.

\section{Cosmology with Inhomogeneous Dark Energy}

\subsection{Cosmological Equations} %for the Dynamics}
The line element of an expanding, spherically symmetric spacetime is given by the Lemaitre-Tolman-Bondi (LTB) metric~\cite{Lemaitre:1933qe,Tolman:1934za,Bondi:1947av}:
\begin{equation}
ds^2 = - dt^2 + X^{2}(r,t)dr^2 + A^{2}(r,t) \left( d\theta^2 +
\sin^2 \theta d\varphi^2 \right)
\label{ltbmet}
\end{equation}
We consider a spherically symmetric energy-momentum tensor:
\be
T_\nu^\mu = \diag(-\rho(r)-\rho_M(r,t),p_r(r),p_t(r),p_t(r))
\label{enmom}
\ee
where $\rho_M(r,t)$ is the matter density and we have allowed for a general inhomogeneous and spherically symmetric static fluid with:
\be
(T_f)_\nu^\mu = \diag(-\rho(r),p_r(r),p_t(r),p_t(r))
\label{enmommon}
\ee
(being $p_t(r)$ the transverse pressure). This expression is motivated by the energy-momentum of a global monopole, which asymptotically is of the form~\cite{Barriola:1989hx}:
\be (T_{\rm mon})_\nu^\mu = \diag(-\eta/r^2,-\eta/r^2,0,0) \label{tmnmonas} \ee
with $\eta$ the symmetry-breaking scale related to the formation of the monopole. The Einstein equations $G_{\mu \nu}=8\pi G T_{\mu \nu}$ may be obtained from equations (\ref{ltbmet}), (\ref{enmom}) and read:

\begin{widetext}
\begin{equation}\label{yht00'}
-2\frac{A''}{AX^2}+2\frac{A'X'}{AX^3}+2\frac{\dot{X}\dot{A}}{AX}+\frac{1}{A^2}+\left(\frac{\dot{A}}{A}\right)^2
-\left(\frac{A'}{AX}\right)^2 = 8 \pi G (\rho_M + \rho)
\end{equation}
\begin{equation}\label{yht10'}
\dot{A}'=A'\frac{\dot{X}}{X}
\end{equation}
\begin{equation}\label{yht11'}
2\frac{\ddot{A}}{A}+\frac{1}{A^2}+\left(\frac{\dot{A}}{A}\right)^2-\left(\frac{A'}{AX}\right)^2
= -8 \pi G p_r
\end{equation}
and
\begin{equation}\label{yht22'}
-\frac{A''}{AX^2}+\frac{\ddot{A}}{A}+\frac{\dot{A}}{A}\frac{\dot{X}}{X}+\frac{A'X'}{AX^3}+\frac{\ddot{X}}{X}
= -8 \pi G p_t\,.
\end{equation}
\end{widetext}
The primes and dots denote derivatives with respect to $r$ and $t$, respectively.
Eq.~(\ref{yht10'}) leads to:
\begin{equation}\label{xarel}
X(r,t)=C(r)A'(r,t)\equiv \frac{A'(r,t)}{\sqrt{1-k(r)}}
\end{equation}
where $C(r)$ is a constant of integration depending only on the coordinate $r$ and $k(r)<1$ is a function associated with the spatial curvature. Thus, the LTB metric (\ref{ltbmet}) takes the form:
\begin{equation}\label{ltbmet2}
ds^2 = - dt^2 + \frac{(A'(r,t))^2}{1-k(r)}dr^2 + A^{2}(r,t) \left(
d\theta^2 + \sin^2 \theta d\varphi^2 \right)\,.
\end{equation}
Taking (\ref{ltbmet2}) to (\ref{yht11'}) we obtain:
\be
\label{freq1}
\dot{A}^2 + 2 A \ddot{A} + k(r) =- 8 \pi G p_r(r) A^2\,,
\ee
which implies accelerating expansion for negative pressure and proper curvature profile.
It is straightforward to check that this equation has the following first integral:
\begin{equation}\label{freq2}
\frac{\dot{A}^2}{A^2} = \frac{F(r)}{A^3} - \frac{8 \pi G}{3}
p_r(r) - \frac{k(r)}{A^2}\,.
\end{equation}
This last equation may also be written as
%\begin{widetext}
\begin{equation}\label{freq3}
H^2(r,t) = H_0^2(r) \left[ \Omega_M(r) \left(\frac{A_0}{A} \right)^3
+ \Omega_X(r) + \Omega_c(r) \left(\frac{A_0}{A} \right)^2
\right]
\end{equation}
%\end{widetext}
provided we define:
\begin{equation}\label{hupple}
H(r,t) \equiv \frac{\dot{A}(r,t)}{A(r,t)}
\end{equation}
\begin{equation}\label{isof}
F(r) \equiv H_0^2(r) \Omega_M(r) A_0^3(r)~,
\end{equation}
where $A_0(r)\equiv A(r,t_0)$ ($t_0$ is the present time), $H_0(r) \equiv H(r,t_0)$,
$\Omega_X (r) \equiv -{8 \pi G p_r(r)}/{3 H_0^2(r)}$ (which is positive for negative radial pressure) and $\Omega_c(r) \equiv 1- \Omega_X (r)-\Omega_M(r)$. Note that the profile of the inhomogeneous dark energy pressure, $\Omega_X(r)$, is in principle arbitrary, being determined by the physical mechanism that induced the inhomogeneous overdensity.

The conservation of the energy-momentum tensor (\ref{enmommon}) leads to the equations:
\ba
\label{conenmomr}
{\dot \rho} + (\rho +p_r)\frac{\dot A'}{A'} + (\rho+p_t)\frac{2{\dot A}}{A}&=& 0\,,\\
p_r' + \frac{2 A'}{A}(p_r-p_t)&=& 0\,.\label{conenmomp}
\ea
In the special case of isotropic pressure ($p_r=p_t\neq 0$) and $A(r,t)=a(t)r$ we obtain the homogeneous dark energy model which has been well studied. In what follows we will assume the presence of anisotropic pressure.

\subsection{%Topological Quintessence from Global Monopoles: A physical model for Hubble-scale Dark energy Inhomogeneities}
A Physical Model: Topological Quintessence from Global Monopoles}
A physical model that can give rise to an energy-momentum tensor of the form (\ref{enmommon}) is a global monopole with a Hubble-scale core embedded in an expanding spacetime. Such a system is described by the action:
\begin{equation}\label{gmaction}
  S=\int d^4 x \sqrt{-g} \left[\frac{m_{Pl}^{~2}}{16\pi}{\cal R}
     -\frac12(\partial_{\mu}\Phi^a)^2-V(\Phi)\right]\,,
\end{equation}
where the Higgs field has three components $\Phi^a ~(a=1,2,3)$, and the potential is
\begin{equation}\label{pote}
V(\Phi)= {1\over 4}\lambda(\Phi^2-\eta^2)^2, ~~
\Phi\equiv\sqrt{\Phi^a\Phi^a}\,,
\end{equation}
with $\eta$ the vacuum expectation value and $\lambda$ a
coupling constant.
In this model, the global monopole field configuration
\begin{equation}\label{hedgehog}
\Phi^a=\Phi(t,r)\hat r^a\equiv
\Phi(t,r)(\sin\theta\cos\varphi,\sin\theta\sin\varphi,\cos\theta)
\end{equation}
is topologically stable and cannot decay to the true vacuum $\Phi =\eta$ in a continuous manner~\cite{Barriola:1989hx}. These field configurations are expected to form during cosmological phase transitions when the vacuum manifold of a field theory has the topology of the sphere $S^2$ via the Kibble mechanism~\cite{Kibble:1980mv}. They are predicted to form initial isocurvature perturbations~\cite{Battye:1998xe}. The monopole solution has localized energy in a core where $\Phi \rightarrow0$ with thickness $d\sim \lambda^{-1/2} \eta^{-1}$ and $\Phi$ reaches asymptotically the vacuum expectation value $\eta$ as a power law. The asymptotic form of the energy-momentum tensor is given by Eq.~(\ref{tmnmonas}). A recently formed global monopole would naturally have a core size comparable to the present Hubble scale

\be \lambda^{-1/2} \eta^{-1}\simeq H_0^{-1} \simeq (10^{-33} eV)^{-1} \label{monscale} \ee

For $\lambda = O(1)$ we obtain $\eta \simeq 10^{-33}eV$, but this energy scale can increase significantly if we allow for small enough values of the coupling constant $\lambda$. Such a field configuration could also be considered as a mechanism leading to spatial variation of fundamental constants (like the fine structure constant) through coupling with the spatially varying scalar field of the monopole. In fact there have been recent claims about detection~\cite{Webb:2010hc} of spatial variation of the fine structure constant along a particular direction in the sky.

By varying the action (\ref{gmaction}) with respect to the metric and with respect to the scalar field, we obtain the following dynamical equations:%for gravity and for the scalar field as
\be
G_{\mu\nu}\equiv{\cal R}_{\mu\nu}-\frac12g_{\mu\nu}{\cal R}={8\pi\over
m_{Pl}^{~2}}T_{\mu\nu},
\ee
\begin{equation}\label{scalar}
\kern1pt\vbox{\hrule height 1.2pt\hbox{\vrule width1.2pt\hskip 3pt
\vbox{\vskip6pt}\hskip 3pt\vrule width 0.6pt}\hrule height
0.6pt}\kern1pt \Phi^a=\frac{\partial V(\Phi)}{\partial\Phi^a}\,,
\end{equation}
where
\begin{equation}\label{ein}
T_{\mu\nu}=\partial_{\mu}\Phi^a\partial_{\nu}\Phi^a
-g_{\mu\nu}\Bigl[\frac12(\partial_{\sigma}\Phi^a)^2+V(\Phi)\Bigr],
\end{equation}
is the energy-momentum tensor of the global monopole. For the metric (\ref{ltbmet2}), its components are
%\begin{widetext}
\be
\begin{aligned}
T_0^0\;=\; &-\rho(r)= -\Bigl({\dot\Phi^2\over2}+{\Phi'^2\over2X(r,t)^2}+{\Phi^2\over
A(r,t)^2}+V(\Phi)\Bigr)\\
T_{0r}\;=\; &\dot\Phi\Phi' \\
T_r^r\;=\; &p_r(r)=\Bigl({\dot\Phi^2\over2}+{\Phi'^2\over2X(r,t)^2}-{\Phi^2\over
A(r,t)^2}-V(\Phi)\Bigr) \\
T_\theta^\theta\;=\; &p_t(r)=T_\phi^\phi=\Bigl({\dot\Phi^2\over2}-{\Phi'^2\over2X(r,t)^2}-V(\Phi)\Bigr)
\end{aligned}
\ee
%\end{widetext}
For a static global monopole they asymptotically reduce to the form (\ref{tmnmonas}) and negative pressure is naturally obtained in the core where the potential energy dominates. Such a static solution has been shown to exist in flat spacetimes~\cite{Barriola:1989hx} and also in expanding spacetimes when the core size is smaller than the Hubble scale (Fig.\,2a of Ref.~\cite{Sakai:1995nh}). Monopoles with cores larger than a critical fraction (of $O(1)$) of the Hubble scale have been shown~\cite{Vilenkin:1994pv,Sakai:1995nh} to become non-static, as their core starts expanding along with the background Hubble expansion. Such an expansion may also be induced by the false vacuum energy of the monopole core leading to {\it topological inflation}. Even though there has been no study of the embedded monopole in an expanding background where matter is also present it is conceivable that a similar behavior will also be present in that case i.e. a static solution will be supported for core sizes up to about the Hubble scale and for a larger size, the core will start expanding along with the background especially when the monopole energy density of the core starts to dominate over the matter energy density. The behavior of such solutions in the presence of matter is an issue under investigation~\cite{inprog}.
\begin{figure}[!t]
\centering
\vspace{0cm}\includegraphics[width=0.48\textwidth]{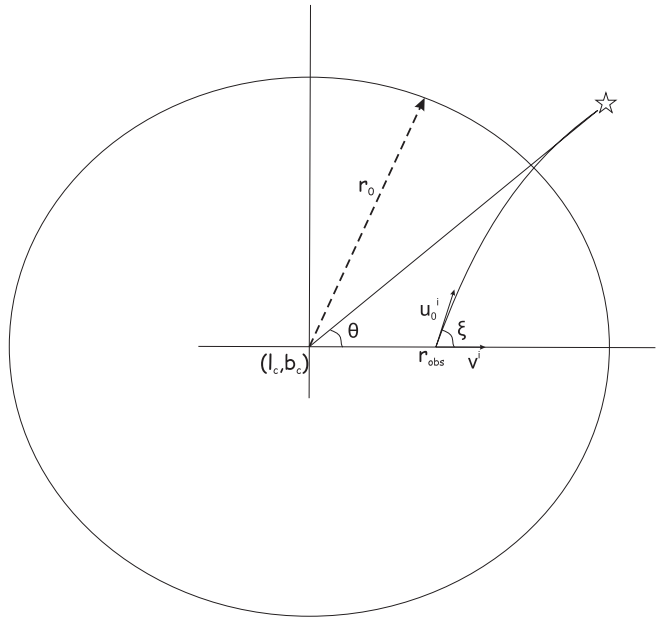}
\caption{The angle $\xi$ between the tangent vector to the geodesic and the unit vector along the axis connecting the center with the observer. \label{xifig} }
\end{figure}

\subsection{General strategy} %the title could maybe be improved, but I think that a section was definitely needed%}
In what follows we focus on the cosmological equation (\ref{freq3}) and use it to derive observational constraints for observers that lie at the center of the inhomogeneity and also for off-center observers. To achieve this goal we follow the following steps:
\begin{enumerate}
\item
 We assume a dark energy profile $\Omega_X(r)$ in Eq.~(\ref{freq3}), characterized by two main parameters: the scale $r_0$ of the inhomogeneity and the magnitude of $\Omega_{X, \text{in}}$ at the center ($\Omega_{X, \text{in}} \equiv \Omega_X(r=0)$).
\item
We solve Eq.~(\ref{freq3}) assuming flatness ($\Omega_M(r)+\Omega_X(r)=1$) with initial conditions taken at the present time $t_0$  as $A(r,t_0)=r$ and evolving backwards in time~\cite{Marra:2010pg}. This initial condition is a gauge choice that corresponds to the usual setting $a(t_0)=1$ made in homogeneous models. Our assumption for flatness is based on the fact that the initial perturbations produced by topological defects are isocurvature while the universe is assumed to be flat before the formation of the Hubble-scale global monopole.
\item
Using the derived solution we solve the geodesic equations and find the light-like geodesics $t(\lambda$), $r(\lambda)$, $\theta(\lambda)$ for on-center and off-center observers. The off-center observer geodesics depend on two additional parameters: the distance $r_{obs}$ of the observer from the center and the angle $\xi$ between the tangent vector to the geodesic and the unit vector along the axis connecting the center with the observer~\cite{Alnes:2006pf}. The value of the angle $\xi$ for each direction in the sky depends on the assumed position (galactic coordinates $(l_c,b_c)$) of the center of the inhomogeneity (i.e. $\xi=\xi(l_c,b_c)$, see Fig.~\ref{xifig}).

\begin{center} \emph{Supernovae} \end{center}

\item
Using the derived geodesics we obtain the luminosity distance and other cosmological observable distances. These distances depend on the model parameters namely $r_0$ and $\Omega_{X, \text{in}} $ for the central observer and in addition to $r_{obs}$ and $(l_c,b_c)$ (the galactic coordinates of the center) for the off-center observer.
\item
We fit the derived forms of the luminosity distance to the measured luminosity distances of the Union2 SnIa data and derive the best-fit values of the parameters $r_0$, $\Omega_{X, \text{in}} $, $r_{obs}$ and $l_c,b_c$ and their $1\sigma$ regions.
\begin{center} \emph{Cosmic Microwave Background} \end{center}
\item
For the derivation of the additional CMB temperature anisotropy induced by the shift of the observer from the center, we use the geodesic equations to derive the dependence of the redshift of the last scattering surface on the angle $\xi$. Having found $z_{ls}(\xi,r_{obs})$ we find the angular dependence of the temperature as
\be T(\xi,r_{obs})=\frac{T_{ls}}{1+z_{ls}(\xi,r_{obs})} \label{tempprof} 
\ee
where $T_{ls}$ is the observed temperature of the last scattering surface. This is assumed to be uniform since we are focusing on the {\it additional} temperature fluctuations induced by the displacement of the observer. We then convert it to temperature fluctuations from the mean and expand in spherical harmonics to find the multipole coefficients.
Comparison with the measured values of the moments (particularly of the dipole) leads to constraints on $r_{obs}$.
\end{enumerate}

\subsection{Evolution of the Scale-Factor $A(r,t)$}
We will consider the following form for the profile of the inhomogeneity
\begin{widetext}
\ba \label{omegam}
   \Omega_M(r)&=&\Omega_{M, \text{out}}+(\Omega_{M, \text{in}}-\Omega_{M, \text{out}}) \, \frac{1-\tanh [(r-r_0) / 2\Delta r ]}{1+\tanh (r_0/ 2\Delta r )} \\
 \label{omegax}
   \Omega_X(r)&=&\Omega_{X, \text{out}}+(\Omega_{X, \text{in}}-\Omega_{X, \text{out}}) \, \frac{1-\tanh [(r-r_0) / 2\Delta r ]}{1+\tanh (r_0/ 2\Delta r )} \,,
\ea
\end{widetext}
Keeping in mind the anticipated physical model of a global monopole, we set $\Omega_{M, \text{out}}=1$, $\Omega_{X, \text{out}}=0$ and since we are considering isocurvature inhomogeneities in a flat universe ($\Omega_c(r)=0$), we keep $\Omega_{M, \text{in}}=1-\Omega_{X, \text{in}}$. In our formulae we keep $\Omega_c(r)$ and $k(r)$, even though these quantities turn out to vanish for the particular profiles we have considered. We also fix $\Delta r=0.35 {\rm \,Gpc}$, corresponding to a fairly sharp transition from `in' to `out'. We are thus left with two parameters characterizing the profile: the radial size $r_0$ and the density at the center $\Omega_{X, \text{in}}$.

In order to solve the dynamical equation (\ref{freq3}) we first determine $H_0(r)$ in units of $Gyr^{-1}$, assuming a simultaneous Big-Bang and, as previously mentioned, an age of the universe $t_0=13.7$\,Gyr. Thus we use:
\begin{equation} \label{age}
   H_0(r)=\frac{1}{t_0}\int\limits^1_0 \frac{dx}{\sqrt{\Omega_M(r)x^{-1}+\Omega_X(r)x^2+\Omega_c(r)}}
\end{equation}
which can be easily derived from Eq.~(\ref{freq3}). We then use this last equation and the initial condition $A(r,t_0)=r$ to solve Eq.~(\ref{freq3}) and obtain numerically the form of $A(r,t)$ and its derivatives.

\subsection{Luminosity distance}
\subsubsection{On-center observer}
In the case of the on-center observer, spherical symmetry with respect to the observer is retained and therefore light travels radially towards the observer. This implies that $d\theta = d\phi =0$ along the geodesic. The remaining two light-like geodesics for radially {\it incoming} light rays are of the form~\cite{Enqvist:2006cg}
\begin{equation}\label{dtdz}
\frac{dt}{dz}=-\frac{A'(r,t)}{(1+z)\dot{A}'(r,t)}
\end{equation}
\begin{equation}\label{drdz}
\frac{dr}{dz}= \frac{c\sqrt{1-k(r)}}{(1+z)\dot{A}'(r,t)}\ .
\end{equation}
where $c\simeq 0.3$ is the velocity on light in units of Gpc/Gyr used for consistency in order to obtain the luminosity distance in Gpc.
Equations~(\ref{dtdz}) and (\ref{drdz}) with initial conditions $t(z=0)=t_0$ and $r(z=0)=0$ determine $t(z)$ and
$r(z)$. The angular diameter and luminosity distance are then simply given by:
\begin{eqnarray}
d_{A}(z,r_0,\Omega_{X, \text{in}}) &=& A(r(z),t(z)) \,, \\
d_{L}(z,r_0,\Omega_{X, \text{in}}) &=& (1+z)^{2} \, d_{A}(z) \label{lumidi} \,.
\end{eqnarray}

The corresponding distance modulus, which is the difference between the apparent
magnitude $m$ and the absolute magnitude $M$, is given by
\begin{equation}
\label{distmodonc}
\mu\equiv m-M=5\log _{10}\left(\frac{d_{\rm L}}{1\ \rm {\rm \,Gpc}}\right)+40
\end{equation}
%\be
%m(z,r_0,\Omega_{M, \text{in}} )=5Log_{10}(10^8 d_L(z,r_0,\Omega_{M, \text{in}} )) \label{apmag}
%\ee
since $d_L$ is obtained in ${\rm \,Gpc}$. It is now straightforward to fit the predicted distance moduli to the observed ones from the Union2 dataset and derive the best-fit values of parameters for the on-center observer.

\subsubsection{Off-center observer}
The light-like geodesics that go through the displaced off-center observer $(t(\lambda),r(\lambda),\theta(\lambda),\phi(\lambda))$ ($\lambda$ is the affine parameter), are derived by solving the geodesic equations in a coordinate system with origin at the center of the inhomogeneity~\cite{Alnes:2006pf} (see Fig.~\ref{xifig}). Taking into account the axial symmetry around the observer, it is clear that we will have $d\phi=0$ along the geodesic. Moreover, due to the the 4-velocity identity $\frac{dx^\mu}{d\lambda}\frac{dx_\mu}{d\lambda}\equiv u^{\mu}u_{\mu}=0$ (valid for light-like geodesics), the time geodesic reduces to a first order ordinary differential equation. The geodesic equation for $\theta(\lambda)$ also reduces to first order thanks to the spherical symmetry of the inhomogeneity (conservation of angular momentum). Finally, the geodesic for $r(\lambda)$ is second order, but it can be split in two first order equations by means of the convenient definition $p\equiv \frac{dr}{d\lambda}$.
The resulting system consists of the following five first order equations~\cite{Alnes:2006pf,Quercellini:2010zr}:
%\begin{widetext}
\begin{equation}
\label{offctrsyst}
\begin{aligned}
    \frac{d t}{d\lambda}\;=\; & -\sqrt{\frac{(A')^{2}}{1-k(r)}\, p^{2} + \frac{J^{2}}{A^{2}}}\,,\\
    \frac{d r}{d \lambda} \;=\; &  c \; p\,,\\
    \frac{d\theta}{d\lambda} \;=\; & c\; \frac{J}{A^{2}}\,, \\
    \frac{d z}{d\lambda} \;=\; & \frac{(1+z)}{\sqrt{\frac{(A')^{2}}{1-k(r)}\, p^{2} + \frac{J^{2}}{R^{2}}}} \left[\frac{A'\dot{A}'}{1-k(r)}\, p^{2} + \frac{\dot{A}}{A^{3}}J^{2} \right],\\
    \frac{d p}{d\lambda} \;=\; & 2\dot{A}'\, p\,\sqrt{\frac{p^{2}}{1-k(r)}\, + \frac{J^{2}}{A^{2}\, A'^{2}}}\\
				&+\,c\,\frac{1-k(r)}{A^{3}A'}J^{2}- c\, \left[\frac{k(r)'}{2-2k(r)} + \frac{A''}{A'}\right] p^{2}\,,
\end{aligned}
\end{equation}
%\end{widetext}
where
\begin{equation}\label{eq:J}
    J\equiv A^{2} \frac{d\theta}{d\lambda} = \textrm{\emph{const}}=J_{0}\,.
\end{equation}
is the conserved angular momentum (whose conservation emerges from the geodesic of $\theta(\lambda)$). The initial conditions are specified at the time $t_0$
when the  photon arrives at the observer's position, which is given by $r=r_{obs}$ and $\theta = 0$ since without loss of generality we may assume that the observer is displaced along the $z$ axis. The photon hits the observer at an angle $\xi$
relative to the $z$-axis. According to the metric (\ref{ltbmet2}), the spatial components of the unit vector
along this axis are
\be
v^i = \frac{\sqrt{1-k(r_{obs})}}{A'(r_{obs},t_0)}(1,0,0)
\ee
where the three components are in the $r$, $\theta$ and $\phi$
direction, respectively. The unit tangent vector $u^i$ to the photon path
at the present time $t_0$ is given by
\be
{\bar u}^i = \left|\frac{d\lambda}{dt}\right|
\left(\frac{dr}{d\lambda},\frac{d\theta}{d\lambda},
  \frac{d\phi}{d\lambda}\right) = -\frac{1}{u}(p,J/A^2,0)
  \label{barui}
\ee
where $u \equiv dt/d\lambda$ and the first factor ensures normalization, $g_{ij}{\bar u}^i {\bar u}^j = 1$ (notice that $\left|\frac{dt}{d\lambda}\right|=g_{ij} u^i u^j$ in view of the light-like geodesic condition $u^{\mu}u_{\mu}=0$). The minus sign in the above equation emerges because we can choose the affine parameter $\lambda$ such that
$\lambda=0$ when $t=t_0$ and $(\frac{dt}{d\lambda})_0=u_0=u(\lambda=0)=-1$ since the time decreases from $t_0$ as $\lambda$ increases.

The angle $\xi$ is then given by:
\be
\label{cosxi}
\cos \xi = g_{ij}{\bar u}^i v^j = -\frac{A'(r_{obs},t_0)}{\sqrt{1-k(r_{obs})}}\frac{p(\lambda=0)}{u_0}\,.
\ee
Notice that at a time $t\neq t_0$ the angle $\xi$ does not represent the angle between the geodesic tangent and the $z$ axis but the angle between the geodesic tangent and the radial unit vector (which coincides with the $z$ axis unit vector only at $t=t_0$ when $\theta=0$). In what follows we refer to $\xi$ as the value obtained from Eq.~(\ref{cosxi}) at the present time $t_0$.

We can now express the initial condition for $p(\lambda)$ in terms of the incident photon angle $\xi$ as
\be
p_0=\frac{\sqrt{1-k(r_{obs})}}{A'(r_{obs},t_0)}\cos \xi = \cos\xi
\label{p0}
\ee

Also using the fact that ${\bar u}^i$ evaluated at $t=t_0$ is a unit vector, we find using the metric (\ref{ltbmet2}) and Eq.~(\ref{barui})
\be
J = A(r_{obs},t_0)\sin\xi = r_{obs} \sin \xi
\label{j0val}
\ee

Thus, once the incident angle $\xi$ and observer shift $r_{obs}$ are specified, the initial conditions to the system (\ref{offctrsyst}) are fixed as follows
\begin{equation}
\label{offctrinit}
\begin{aligned}
    t(0)\;=\; &t_0\\
    r(0)\;=\; &r_{obs}\\
    \theta(0)\;=\; & 0 \\
    z(0)\;=\; & 0\\
    p(0)\;=\; & p_0
\end{aligned}
\end{equation}
Thus, the numerical solution of the system (\ref{offctrsyst}) with initial conditions (\ref{offctrinit}) is straightforward and leads to the geodesics $t(\lambda,\xi,r_{obs})$,
$r(\lambda,\xi,r_{obs})$, $\theta(\lambda,\xi,r_{obs})$ and $z(\lambda,\xi,r_{obs})$. By inverting the expression $z=z(\lambda,\xi,r_{obs})$ the affine parameter $\lambda$ may be replaced by the redshift in $t$, $r$ and $\theta$.

The luminosity distance
is now obtained from the geodesics for the case of an off-center observer as~\cite{Blomqvist:2009ps,Alnes:2006uk}
\begin{widetext}
\begin{equation}
\label{lumdistoff}
d_{\rm L}^4(z,\xi,r_{obs})=(1+z)^8 \frac{A(r(z...),t(z...))^4\sin^2\theta(z...)}{\sin^2\xi}\Bigg[ \frac{A'(r(z...),t(z...))^2}{A(r(z...),t(z...))^2(1-k(r(z...)))}
\Bigg( \frac{\partial r(z...)}{\partial \xi} \Bigg)^2+\Bigg( \frac{\partial \theta(z...)}
{\partial \xi} \Bigg)^2 \Bigg]\ ,
\end{equation}
\end{widetext}
where $...\equiv ,\xi,r_{obs}$ and the partial derivatives with respect to $\xi$ are obtained numerically by evaluating the geodesics at slightly different values of $\xi$. It is easy to see that for the on-center observer we have $\frac{\partial r(z,\xi,r_{obs}=0)}{\partial \xi}=0$ and $\frac{\partial \theta(z,\xi,r_{obs}=0)}{\partial \xi}=1$ (since $\theta=\xi$) and therefore we re-obtain the expression of the luminosity distance for the on center observer (\ref{lumidi}).

In order to compare the predicted luminosity distance with the corresponding observed one through SnIa data, it is important to derive the dependence of the angle $\xi$ on the coordinates of each SnIa. Let us assume an equatorial coordinate
system, where coordinates are given by the right ascension $\alpha$ and declination
$\delta$, since the directions of the SnIa of the Union2 dataset are provided in this coordinate system~\cite{Blomqvist:2010ky}. Being the angle between the center-observer axis and the observer-SnIa direction, $\xi$ depends on both the coordinates of the inhomogeneity center $(\alpha_c,\delta_c)$ and on the coordinates of each SnIa $(\alpha,\delta)$. It may be shown~\cite{Blomqvist:2009ps} that the function $\xi(\alpha,\delta,\alpha_c,\delta_c)$ is given by
\begin{widetext}
\begin{equation}
\label{xiacdccon}
\xi =\arccos\left[-\sin(90^\circ - \delta_{\rm
c}) \sin(90^\circ-\delta)\cos(\alpha+(360^\circ - \alpha_{\rm c}))+
\cos(90^\circ - \delta_{\rm
c}) \cos(90^\circ-\delta)\ \right],
\end{equation}
\end{widetext}
Thus, equation (\ref{xiacdccon}) translates the parameter $\xi$ to the coordinates of the center of the inhomogeneity $(\alpha_c,\delta_c)$ and introduces the SnIa equatorial coordinates $(\alpha,\delta)$. This allows the fit to the SnIa Union2 data and the derivation of the best fit parameter values $(r_{obs},\alpha_c,\delta_c)$.

\subsection{CMB anisotropies for an off-center Observer}
We assume isotropic temperature on the last scattering surface and focus only on those additional temperature anisotropies that emerge due to the shift of the observer from the center of the dark energy inhomogeneity. Thus the CMB temperature measured by the off-center observer is given by Eq.~(\ref{tempprof}).
%\be
%\label{tempofxi}
%T(\xi,r_{obs})=\frac{T_{ls}}{1+z_{ls}(\xi,r_{obs})}\,,
%\ee
%where $T_{ls}$ is the temperature at the last-scattering surface, $\xi$ is the angle defined in the previous subsection and $z_{ls}(\xi,r_{obs})$ is the redshift corresponding to the %last-scattering surface at the direction corresponding to $\xi$.
In order to determine $z_{ls}(\xi,r_{obs})$ we proceed as follows:
\begin{itemize}
\item
We find the light-like geodesics for an on-center observer with fixed $r_0$ and $\Omega_{X, \text{in}}$ and determine the time at decoupling $t_{ls}=t(\lambda_{ls})$ where $\lambda_{ls}$ is determined by solving the equation $z(\lambda_{ls})=1100$.
\item
For a given off-center observer shifted by $r_{obs}$ and a given direction $\xi$ we find the light-like geodesics and determine $\lambda_{ls}(r_{obs},\xi)$ by solving the equation $t_{ls}=t(\lambda_{ls},r_{obs},\xi)$ where $t_{ls}$ is the decoupling time as determined for the on-center observer.
\item
We determine $z_{ls}(\xi,r_{obs})\equiv z(\lambda_{ls},r_{obs},\xi)$ where $z(\lambda_{ls},r_{obs},\xi)$ is the redshift along the geodesic as determined in the previous subsection.
\end{itemize}
In order to calculate the predicted temperature fluctuations we must also obtain the predicted mean value $\bar T$ of the CMB temperature $T(\xi,r_{obs})$. Notice that this value is not exactly equal to the temperature $T_{ls}$ at the last scattering surface. The mean temperature $\bar T$ is evaluated using Eq.~(\ref{tempprof}) as~\cite{Alnes:2006pf}
%\begin{widetext}
\ba
\label{meantemp}
{\bar T}(r_{obs})\equiv\frac{1}{4\pi}\int\!d\Omega\, T(\xi,r_{obs}) =
\frac{T_{ls}}{2}\int_0^\pi d\xi\frac{\sin{\xi}}{1+z_{ls}(\xi,r_{obs})}\nonumber\\[2mm]
\ea
%\end{widetext}
The predicted angular dependence of the temperature fluctuations may now be easily evaluated using Eq.~(\ref{meantemp}) as:
%\begin{widetext}
\ba
\label{tempfluct}
\frac{\Delta T}{{\bar T}}(\xi,r_{obs})&\equiv&\frac{T(\xi,r_{obs})-
  {\bar T}(r_{obs})}{{\bar T}(r_{obs})}\nonumber\\&=&\frac{{\bar z}(r_{obs})-z_{ls}(\xi,r_{obs})}{1+z_{ls}(\xi,r_{obs})}
\ea
%\end{widetext}
where ${\bar z}(r_{obs})$ is defined as
\be
\label{barz}
{\bar z}(r_{obs})\equiv 2\left[\int_0^\pi\!d\xi
\frac{\sin{\xi}}{1+z(\xi,r_{obs})}\right]^{-1} -1
\ee
The predicted values of the multipole coefficients are obtained by expanding the temperature fluctuations into spherical harmonics as
\be
\label{dtexpy}
\frac{\Delta T}{{\bar T}}(\xi,r_{obs}) = \sum_{l,m}a_{lm}Y_{lm}(\xi,\phi)
\ee
where the values of the multipole moments $a_{lm}$ are obtained as
\be
\label{cmbalm}
a_{lm}=
\int_0^{2\pi}\!\int_0^\pi\!\frac{\Delta T}{{\bar T}}(\xi,r_{obs})\,Y_{lm}^*(\xi,\phi) \sin{\xi}\, d\xi\,
d\phi
\ee
Due to axial symmetry of the temperature perturbations, all the $a_{lm}$ will vanish, except those with $m=0$. The observed  CMB dipole $a_{10}$ is of the order $10^{-3}$. By demanding that the predicted value of $a_{10}$ does not exceed $10^{-3}$ we may obtain an upper bound on $r_{obs}$.

There is a Newtonian analytical approximation that can be made in order to obtain an expression for $z(\xi,r_{obs})$ valid for small observer displacement $r_{obs}$. The cosmological setup corresponding to a displaced observer from the center of a spherical inhomogeneity may be replaced~\cite{Tomita:1999rw,Alnes:2006pf} by that of an observer with a specific peculiar velocity $v_{obs}$. This peculiar velocity is due to  the increased value of the Hubble parameter $H_{in}$ inside the inhomogeneity compared to the corresponding value that would exist if the inhomogeneity was not present ($H_{out}$). It is given by
\be
q=\frac{v_{obs}}{c} = \frac{H_{in}-H_{out}}{c}r_{obs} \label{qdef}
\ee
For example for the profiles of Eqs.~(\ref{omegam}), (\ref{omegax}) with $r_0=3.37Gpc$, $\Omega_{X, \text{in}}=0.69$ we have \be q\simeq 0.07 \frac{r_{obs}}{1\;Gpc} \label{qval} \ee 

This peculiar velocity leads to the observation of Doppler shifted photon frequencies and to an anisotropic temperature profile $T_{obs}(\xi,r_{obs})$. The ratio of this anisotropic temperature over the isotropic temperature $T_c$ observed by the central observer is~\cite{Alnes:2006pf} 
\be
\frac{T_{obs}}{T_c}=\frac{\sqrt{1-q^2}}{1-q \cos{\xi}}=\frac{1+z_{ls,c}}{1+z_{ls}(\xi,r_{obs})}\simeq \frac{z_{ls,c}}{z_{ls}(\xi,r_{obs})} \label{temprat}
\ee
where the first equality is obtained by a Doppler shift analysis and $z_{ls,c}\simeq 1100$ is the redshift of the last scattering surface as detected by the central observer. For small $q$ we find
\be
z_{ls}(\xi,r_{obs})\simeq z_{ls,c} (1-q \cos{\xi}) \label{zlsanal}
\ee
Thus, using Eq.~(\ref{zlsanal}) in Eq.~(\ref{tempprof}) we have an approximate analytic expression for the predicted temperature profile which may lead to the evaluation of the predicted low multipole moments. We have checked that our numerical results presented in the next section are in rough agreement with the predictions of this Newtonian analytical model. This model correctly predicts the qualitative angular dependence of $z_{ls}$ on $\xi$ ($\propto 1-q\cos\xi$) but we found it to be off by a factor $\sim 2$ on the amplitude $q$ for the models considered. 

\section{Numerical Analysis: Observational Constraints}

In this section, we implement the methods described in the previous section to impose observational constraints on the parameters of the model. In order to do so we will use the Union2 SnIa data and also COBE/WMAP results for the observed CMB 
low multipole moments.

\subsection{Constraints from the Union2 SnIa data}

\subsubsection{On-Center Observer}

\begin{figure}[!t]
\centering
\vspace{0cm}\includegraphics[width=0.48\textwidth]{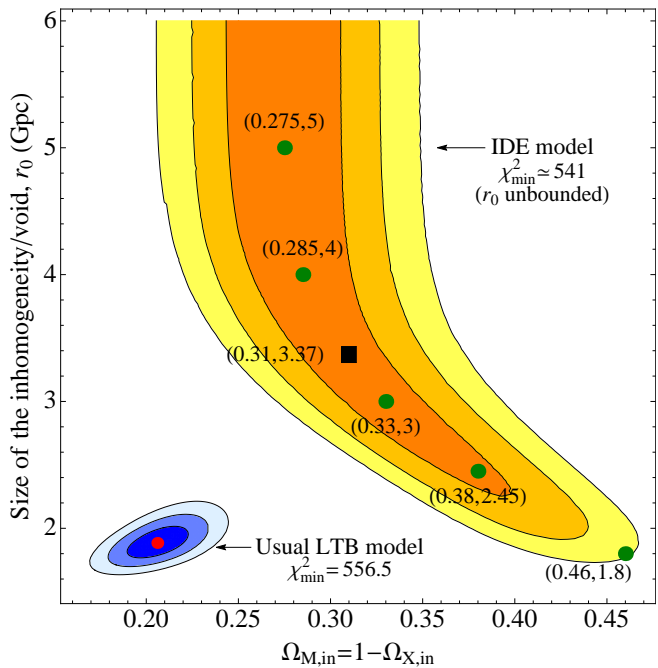}
\caption{The $1\sigma$, $2\sigma$ and $3\sigma$ contours in the parameter space $r_0-\Omega_{M, \text{in}}$ for an on-center observer, obtained from Eq.~(\ref{chi2}). The parameter values $(r_0,\Omega_{X, \text{in}})=(3.37 {\rm \,Gpc},0.69)$ provide a good fit to the data and will be often assumed when dealing with the off-center case. The corresponding $\chi^2$ contours for the usual LTB model (only matter\,+\,curvature, $\Omega_{M,out}=1$) are also shown. Note that the relation $\Omega_{M, \text{in}}=1-\Omega_{X, \text{in}}$ is obviously only valid for the IDE model.\label{ontrfig}}
\end{figure}

\begin{figure*}[!t]
\centering
\rotatebox{0}{\hspace{0cm}\resizebox{0.8\textwidth}{!}{\includegraphics{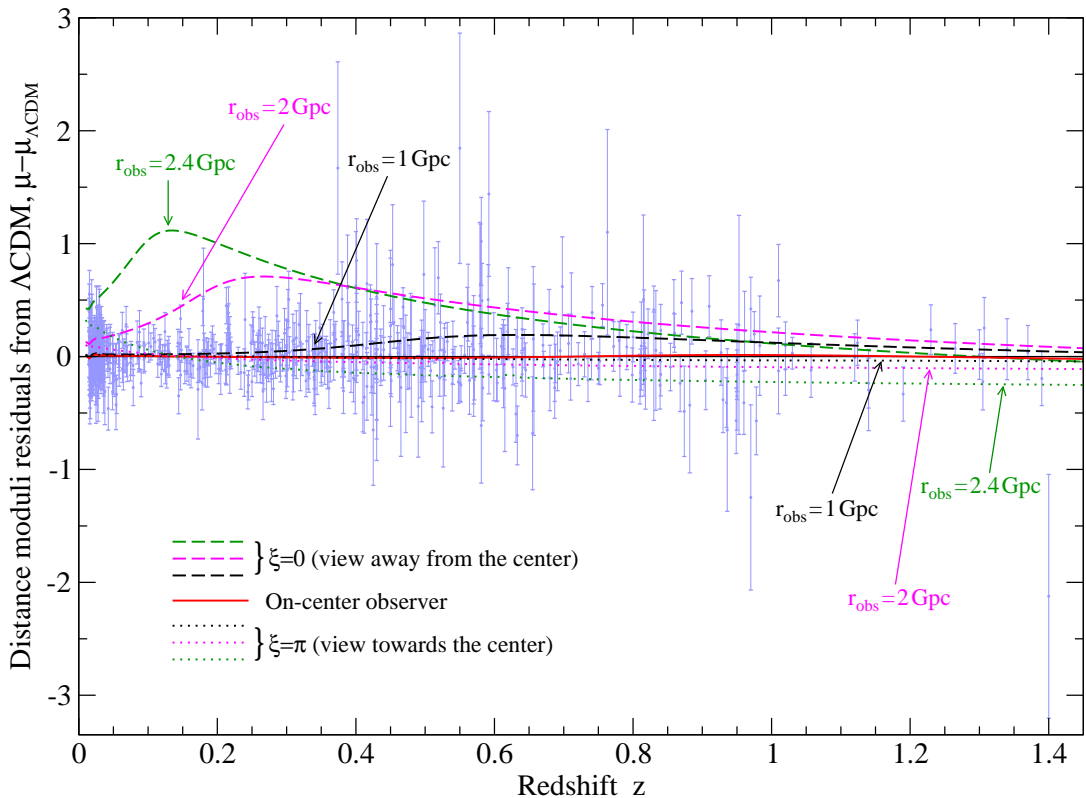}}}
\hspace{0pt}\caption{The distance modulus as obtained from Eqs.~(\ref{lumdistoff}), (\ref{distmodonc}) in the direction of the center ($\xi=\pi$) and in the opposite direction ($\xi=0$), for various values of the observer shift $r_{obs}$ and fixed $(r_0,\Omega_{X, \text{in}})=(3.37 {\rm \,Gpc},0.69)$. The offset $\mu_0$ has been set to the best-fit value corresponding to $\xi=\pi$. \label{distmod} }
\end{figure*}

The Union2 SnIa data points are given, after the corrections have been
implemented, in terms of the distance modulus \be
\mu_\text{obs}(z_i)\equiv m_\text{obs}(z_i) - M \label{mug}.\ee The parameters of the theoretical
model are determined by minimizing the quantity:
%\begin{widetext}
\be
\chi^2_{1} (r_0,\Omega_{X, \text{in}},\mu_0)= \sum_{i=1}^N \frac{(\mu_\text{obs}(z_i) -
\mu_\text{th}(z_i,r_0,\Omega_{X, \text{in}})-\mu_0)^2}{\sigma_{\mu \!\; i}^2 } \label{chi2},
\ee
%\end{widetext}
where
$N=557$ and $\sigma_{\mu \!\; i}^2$ are the errors due to flux
uncertainties, intrinsic dispersion of SnIa absolute magnitudes and
peculiar velocity dispersion. These errors are assumed to be
Gaussian and uncorrelated. The theoretical distance modulus $\mu_{th}(z_i,r_0,\Omega_{X, \text{in}})$ is given by Eq.~(\ref{distmodonc}) and we have allowed for an offset $\mu_0$ which is dealt with either by marginalization or by minimization (the choice does not affect the fit of the other parameters)~\cite{Nesseris:2005ur}.

\begin{figure*}[]
\centering
\rotatebox{0}{\hspace{0cm}\resizebox{0.99\textwidth}{!}{\includegraphics{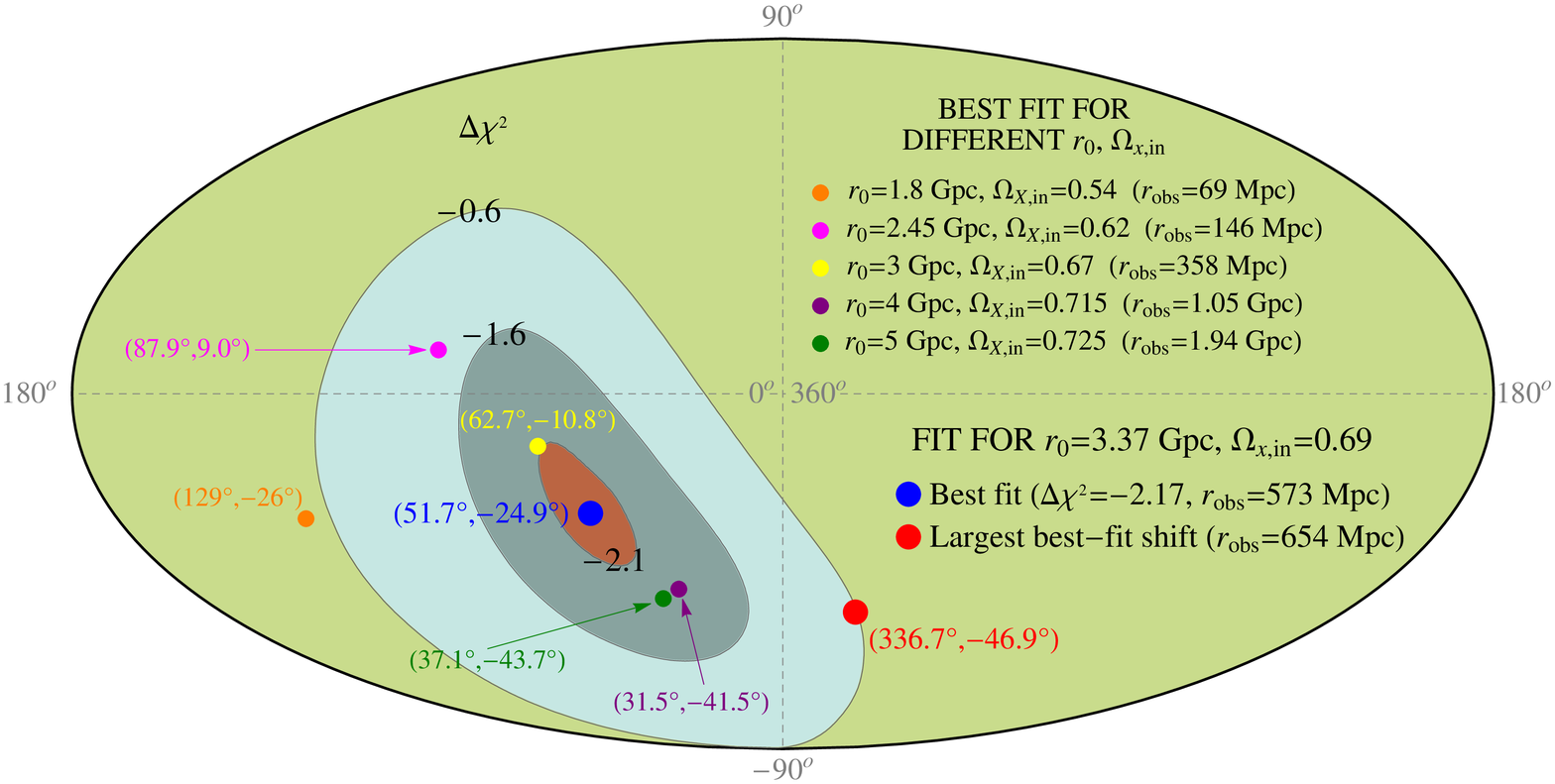}}}
\hspace{0pt}
\caption{Contour plot of $\chi^2_{2min}(l_c,b_c)$ in galactic coordinates obtained after minimization with respect to $r_{obs}$ and $\mu_0$. The $\chi^2$ contours shown correspond to $r_0=3.37$ and $\Omega_{X,\text{in}}=0.69$, and $\Delta\chi^2$ represents the improvement in the fit with respect to the on-center case. The best-fit directions towards the inhomogeneity center for all the $(r_0,\Omega_{X,\text{in}})$ values in Fig.~\ref{ontrfig} are also shown. Finally, the red point indicates the largest best-fit value for the shift of the observer (see the text and Fig.~\ref{fig5}).\label{fig4}}
\end{figure*}

\begin{figure*}[]
\centering
\rotatebox{0}{\hspace{0cm}\resizebox{0.99\textwidth}{!}{\includegraphics{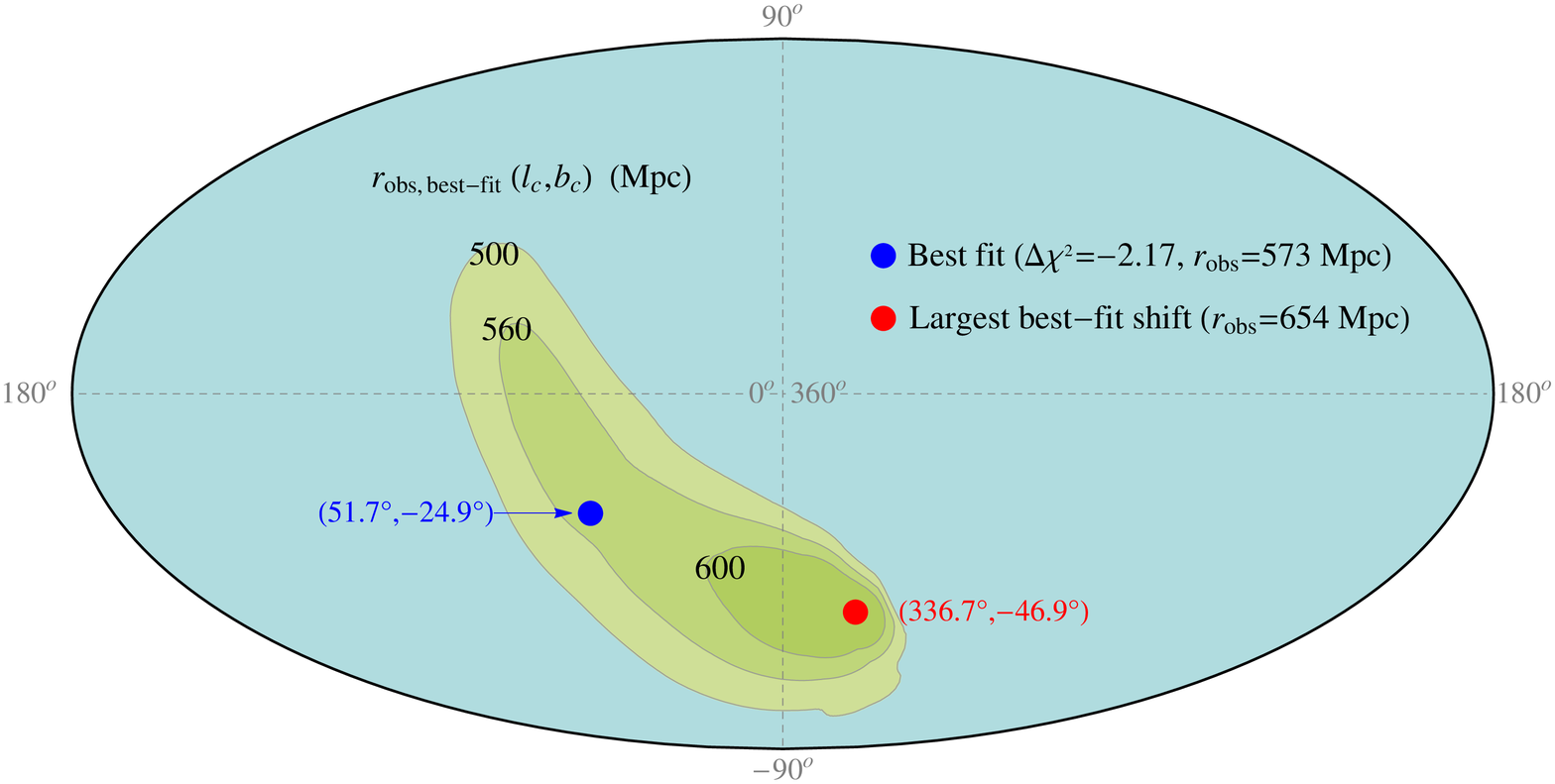}}}
\hspace{0pt}
\caption{Contour plot in galactic coordinates of the best-fit value of $r_{obs}$ as a function of the location of the inhomogeneity center, given by the coordinates $(l_c,b_c)$). The largest best-fit ($r_{obs}=654$~Mpc) and absolute best-fit ($r_{obs}=573$~Mpc, corresponding to the minimum value of $\chi_2^2$ in Eq.~(\ref{chi2off})) values for the observer's shift are also shown. \label{fig5}}
\end{figure*}

In Fig.~\ref{ontrfig} we show the $1\sigma$, $2\sigma$ and $3\sigma$ contours in the parameter space $r_0-\Omega_{M, \text{in}}$ (where $\Omega_{M, \text{in}}=1-\Omega_{X, \text{in}}$) obtained from Eq.~(\ref{chi2}) after minimization with respect to the nuisance parameter $\mu_0$. As could have been anticipated, there is no upper observational limit on the size of the inhomogeneity, since the homogeneous limit $r_0\rightarrow \infty$ corresponds to the \lcdm model, which is indeed consistent with the data. What is more interesting is the fact that the range
%$r_0<1.9 {\rm \,Gpc}$ is excluded by the Union2 SnIa data at the $2\sigma$ level
$r_0\lesssim1.8 {\rm \,Gpc}$ is excluded by the Union2 SnIa data at the $3\sigma$ level. The range of allowed values of $\Omega_{X, \text{in}}$ is %$0.56<\Omega_{X, \text{in}}<0.77$ at the $2\sigma$ level.
$0.53<\Omega_{X, \text{in}}<0.79$ at the $3\sigma$ level. In Fig.~\ref{ontrfig} we also show the corresponding $\chi^2$ contours for the usual LTB model (only matter\,+\,curvature, $\Omega_{M,out}=1$). Notice the reduced size of the allowed region in the parameter space and also the reduced quality of the fit ($\chi^2\simeq556.5$) as compared to the IDE and $\Lambda$CDM models ($\chi^2\simeq541$).

In the remainder of the section we move to investigate the SnIa and CMB results for an off-center observer. Since in that case there are three additional parameters, $(r_{obs},\xi(\alpha_c,\delta_c))$, we will consider a discrete set of values for $(r_0,\Omega_{X, \text{in}})$, as shown in Fig.\,\ref{ontrfig}, with special attention to the values $(3.37 {\rm \,Gpc},0.69)$. These values provide a good fit to the Union2 data for the on-center observer, while being distinct from the $\Lambda$CDM limit.\\

\begin{figure*}[]
\centering
\hspace{-1cm}\includegraphics[width=0.8\textwidth]{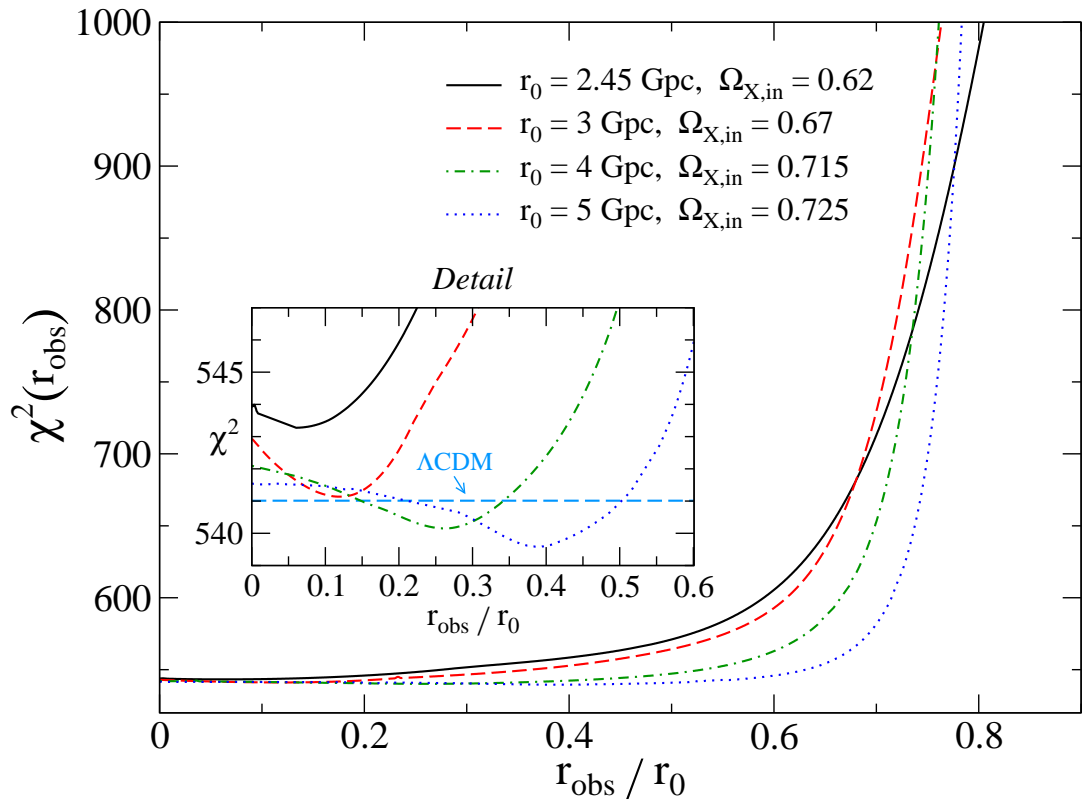}
\caption{The form $\chi^2(r_{obs})$ after minimization with respect to $\alpha_c,\delta_c,\mu_0$ for various values of $r_0$, $\Omega_{X, \text{in}}$. The fit is statistically acceptable out to values of $r_{obs}\simeq 0.7 \, r_0$. Clearly, in this case the Copernican principle is respected. \label{robsmax} }
\end{figure*}

\subsubsection{Off-Center Observer}

In the case of the off-center observer, the theoretical distance moduli depend not only on $(z,r_0,\Omega_{X, \text{in}})$, but also on $(r_{obs},\xi(\alpha_c,\delta_c,\alpha,\delta))$ (see Eqs.~(\ref{distmodonc}), (\ref{lumdistoff}), (\ref{xiacdccon})). In order to show the effect of the additional parameters on the luminosity distance and the corresponding distance modulus, we show in Fig.~\ref{distmod} the distance modulus as obtained from Eqs.~(\ref{distmodonc}), (\ref{lumdistoff}) in the direction of the center ($\xi=\pi$) and in the opposite direction ($\xi=0$), for various values of the observer shift $r_{obs}$ and fixed $(r_0,\Omega_{X, \text{in}})=(3.37 {\rm \,Gpc},0.69)$. We also superpose the Union2 data. Clearly, values of $r_{obs}> 1{\rm \,Gpc}$ in the direction away from the center do not provide a good fit. However, in the direction {\it towards the center}, the fit is good even for  large $r_{obs}$ approaching the size of the inhomogeneity.
\begin{figure*}[!t]
\centering
\rotatebox{0}{\hspace{0cm}\resizebox{0.8\textwidth}{!}{\includegraphics{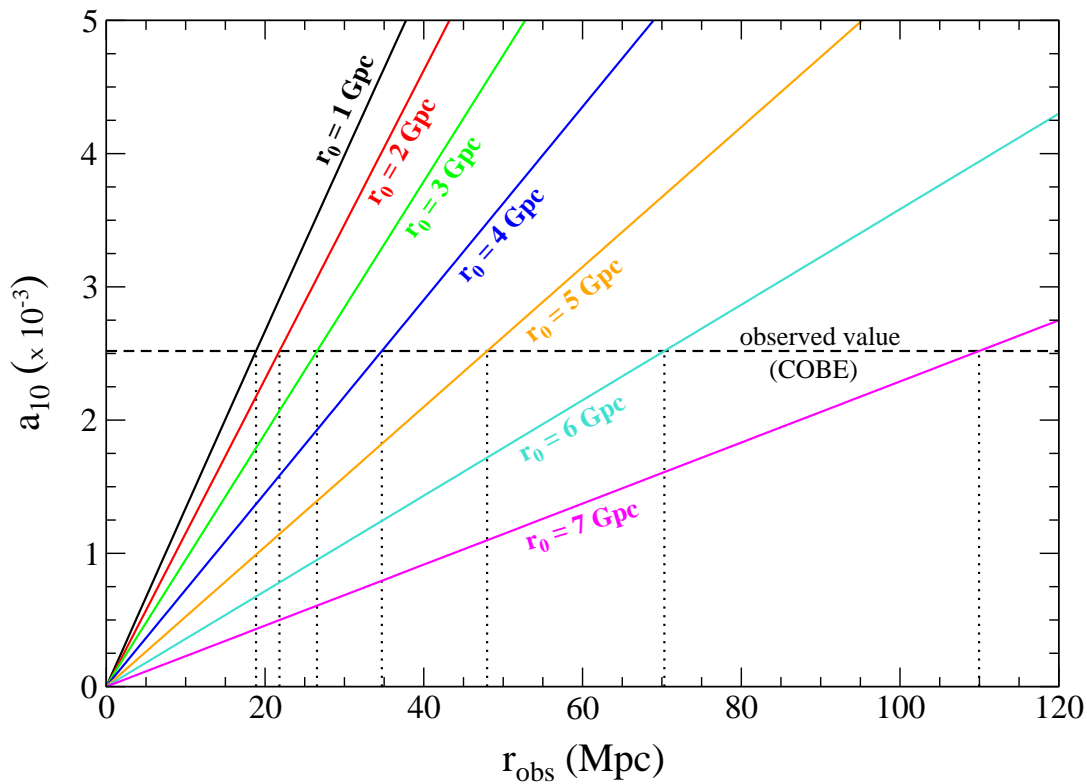}}}
\hspace{0pt}
\caption{The predicted CMB dipole $a_{10}(r_{obs})$ for $\Omega_{X, \text{in}}=0.69$ and various values of the inhomogeneity scale $r_0$. The dashed line corresponds to the observed value (whose uncertainty is too small to be shown, cf. Eq.~\ref{cobedip2})\vspace{2.5mm}\label{fig7}}
\end{figure*}

\begin{figure}[!t]
\centering
\rotatebox{0}{\hspace{0cm}\resizebox{0.48\textwidth}{!}{\includegraphics{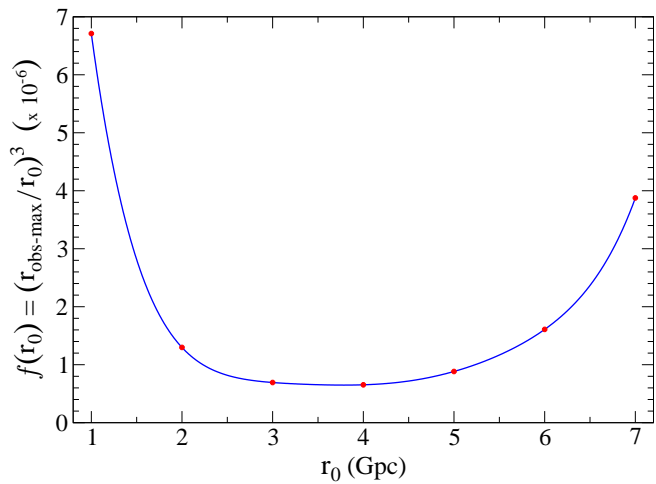}}}
\hspace{0pt}
\caption{The spatial fraction $f(r_0)\equiv(\frac{r_{obs-max}}{r_0})^3$ where the observer needs to be confined in order to be consistent with the value of the observed CMB dipole, for $(r_0,\Omega_{X, \text{in}})=(3.37 {\rm \,Gpc},0.69)$. \label{fig8} }
\end{figure}

\begin{figure*}[!t]
\centering
\rotatebox{0}{\resizebox{0.48\textwidth}{!}{\includegraphics{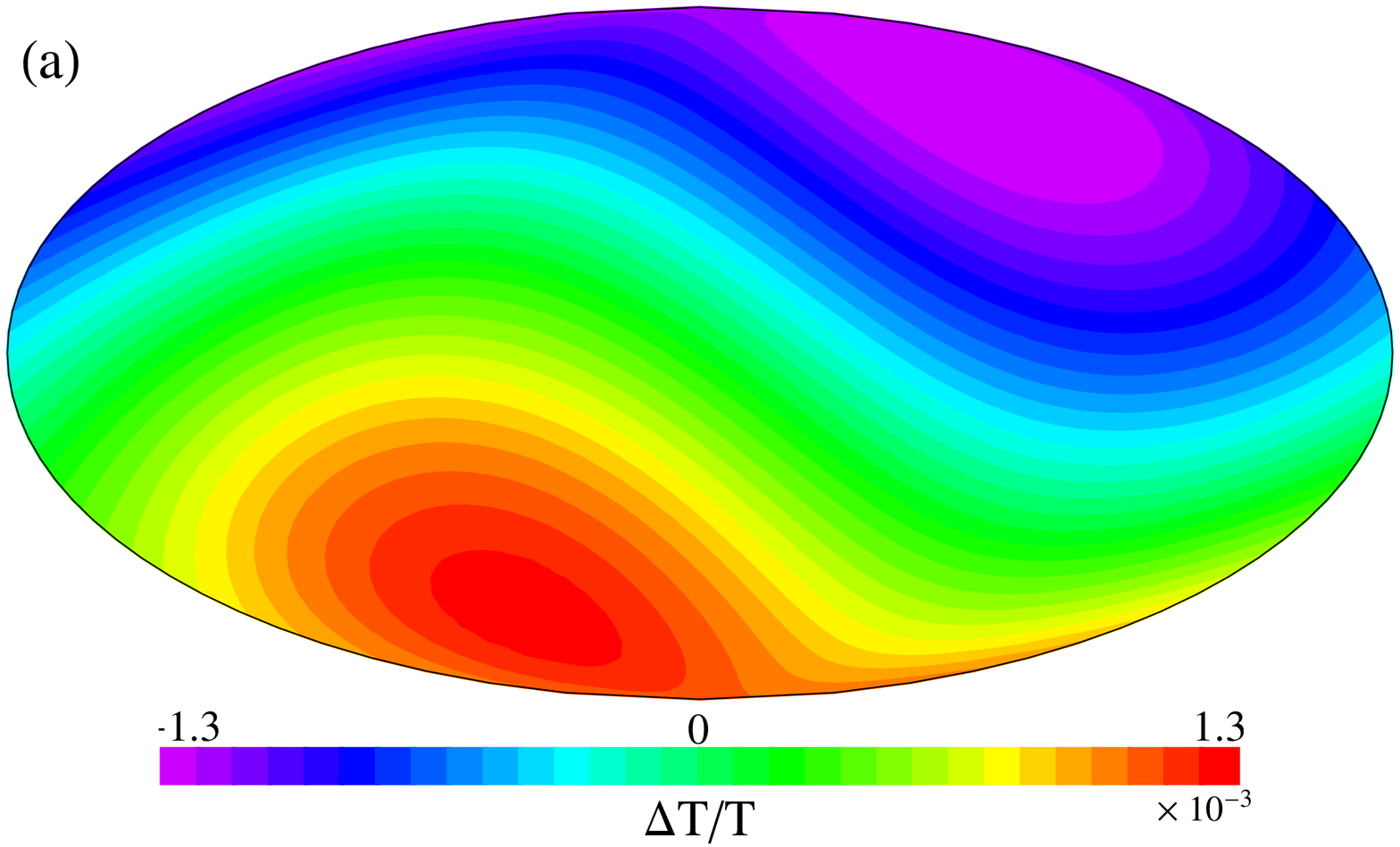}}}
\rotatebox{0}{\resizebox{0.48\textwidth}{!}{\includegraphics{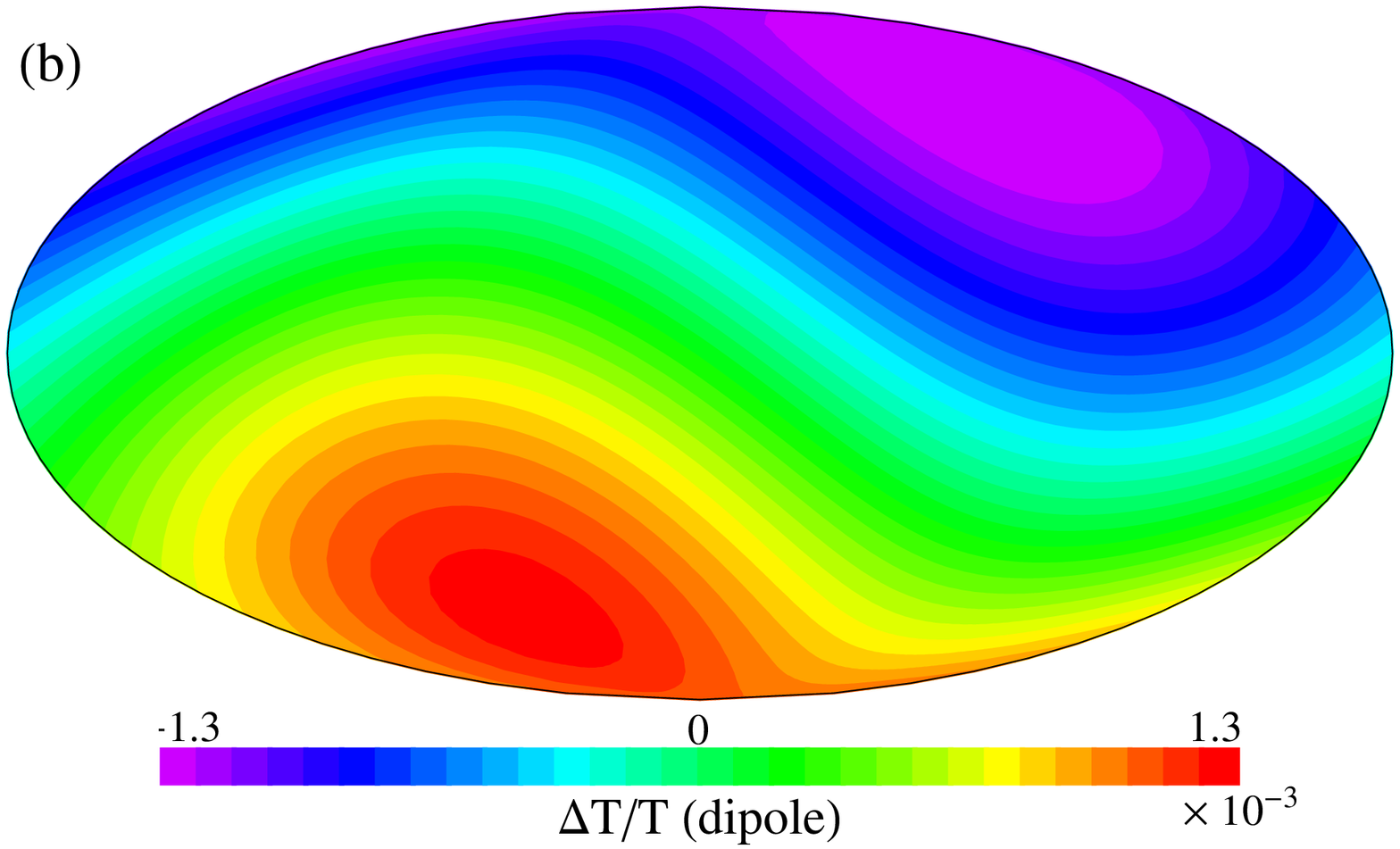}}}\\[5mm]
\rotatebox{0}{\resizebox{0.48\textwidth}{!}{\includegraphics{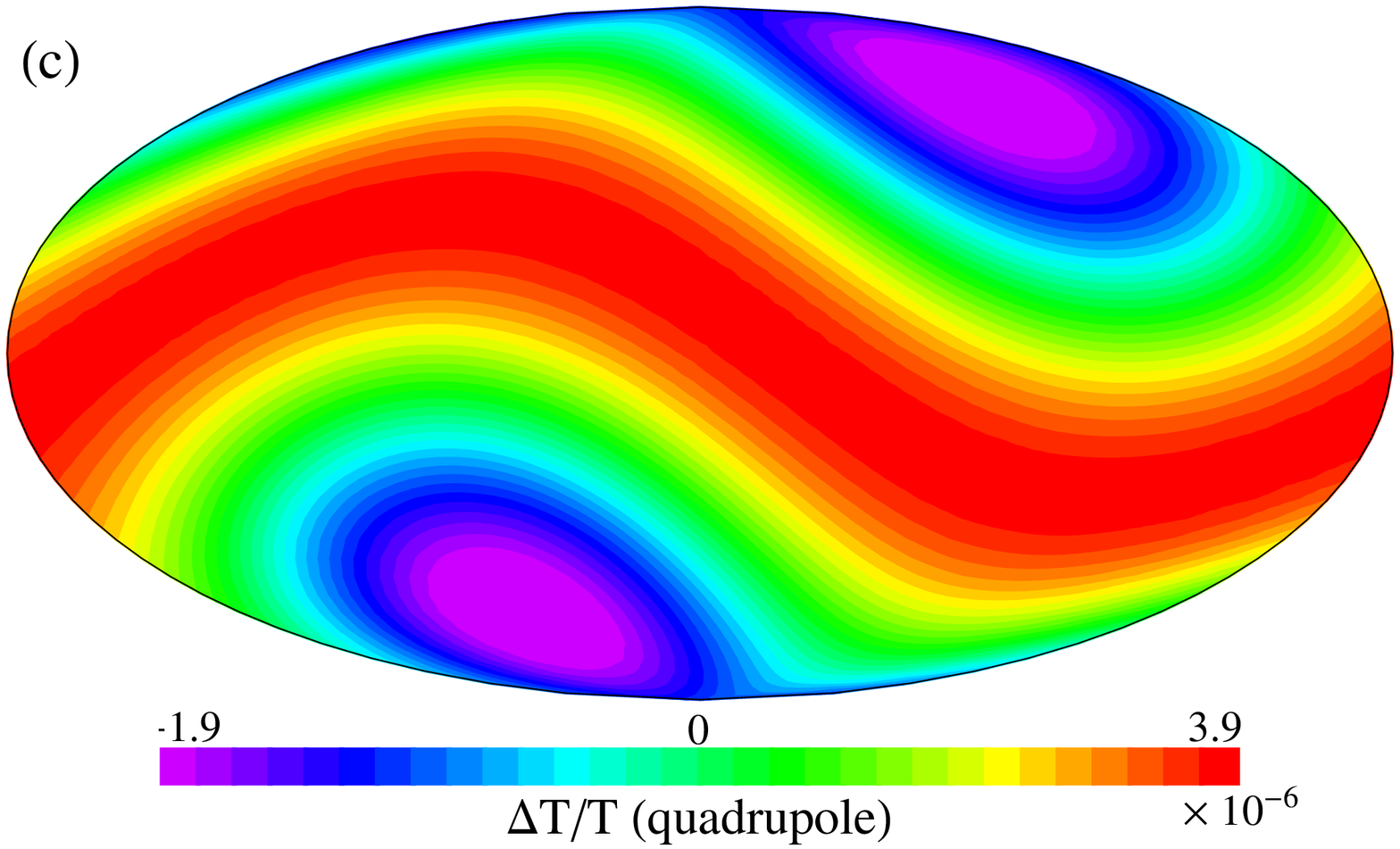}}}
\rotatebox{0}{\resizebox{0.48\textwidth}{!}{\includegraphics{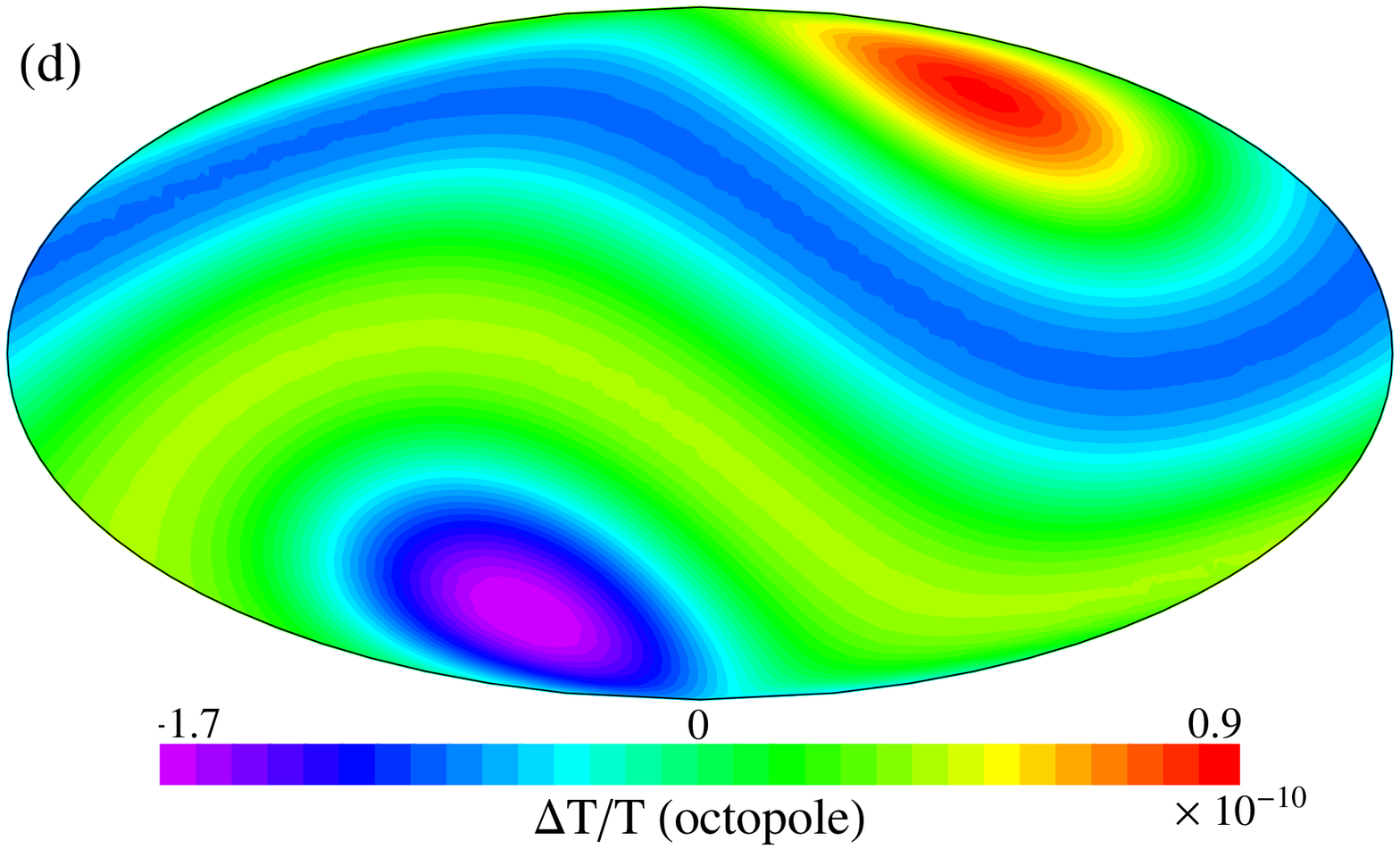}}}
\hspace{0pt}\caption{The predicted CMB maps of additional temperature fluctuations in galactic coordinates corresponding to the full map (a), dipole (b) quadrupole (c) and octopole (d). We have selected the direction of the preferred axis to coincide with the direction of the observed CMB dipole and used $(r_0,\Omega_{X, \text{in}})=(3.37 {\rm \,Gpc},0.69)$, $r_{obs}=30Mpc$. Notice the rapid increase of the maximum allowed $r_{obs-max}$ as we increase the inhomogeneity size $r_0$. It is anticipated that as we reach the value $r_0\simeq13.7$ Gpc (the distance to the last-scattering surface) the model becomes indistinguishable from the $\Lambda$CDM and therefore no constraints are imposed on the value of $r_{obs}$ using the CMB dipole.}\vspace{1cm} \label{fig9}
\end{figure*}

In order to find the best-fit values for the parameters of the model, and assuming that $(r_0,\Omega_{X, \text{in}})$ are fixed, we need to minimize the following $\chi^2$ function:
\begin{widetext}
\be
\chi^2_{2} (r_{obs},\alpha_c,\delta_c,\mu_0)= \sum_{i=1}^N \frac{(\mu_{obs}(z_i,\alpha_i,\delta_i) -
\mu_{th}(z_i,\alpha_i,\delta_i,r_{obs},\alpha_c,\delta_c)-\mu_0)^2}{\sigma_{\mu \; i}^2 } \label{chi2off},
\ee
\end{widetext}
where $\mu_{th}$ is obtained by combining Eqs.~(\ref{distmodonc}) and (\ref{lumdistoff}).
The contour plot of $\chi^2_{2min}(l_c,b_c)$ obtained for $(r_0,\Omega_{X,\text{in}})=(3.37{\rm\,Gpc},0.69)$ after conversion to galactic coordinates and minimization with respect to $r_{obs}$ and $\mu_0$ is shown in Fig.~\ref{fig4}. The best-fit coordinate values for the location of the center of the dark energy spherical inhomogeneity are $(l_c,b_c)=(51.7^\circ\pm 60^\circ,-24.9^\circ\pm 65^\circ)$  where the errors correspond to a rough estimate of the $1\sigma$ region. The corresponding $1\sigma$ range for the observer shift is $0<r_{obs}<573 {\rm \,Mpc}$. 
In Table~\ref{tab1} and in Fig.~\ref{fig4} we also show the best-fit directions towards the inhomogeneity center obtained for the other values $(r_0,\Omega_{X,\text{in}})$ shown in Fig.~\ref{ontrfig}. Even though these directions are in a similar region in the sky, the uncertainty is clearly very large. 

Despite of the existence of three additional parameters $(r_{obs},\alpha_c,\delta_c)$, the improvement of the fit compared to on-center case is marginal (for $(r_0,\Omega_{X,\text{in}})=(3.37{\rm Gpc},0.69)$ we obtain $\Delta \chi^2 = -2.17$ at the best-fit point $(r_{obs},l_c,b_c)=(573{\rm \,Mpc},51.7^\circ,-24.9^\circ)$). The improvement with respect to the \lcdm model is even smaller: we obtain $\Delta \chi^2 \simeq -0.3$ for the same values of $(r_0,\Omega_{X,\text{in}})$.   We conclude that there is no statistically significant advantage of this model over \lcdm. However, the direction of the mildly preferred axis is not far from the direction of the CMB dipole $(l_d,b_d)\simeq (264^\circ,48^\circ)\equiv (84^\circ,-48^\circ)$ where we have allowed for identification of opposite directions.

\begin{table}

{\renewcommand\arraystretch{1.5}\setlength\tabcolsep{0.3cm}\begin{tabular}{|c|c|c|c|c|c|}
\hline
$r_0$ & $\Omega_{X,\text{in}}$ & $l_c$ & $b_c$ & $r_{obs}$ & $\chi^2_{min}$\\
\hline
1.8 & 0.54 & 129$^\circ$ & -26$^\circ$& 0.069&552.4\\
2.45 & 0.62 & 87.9$^\circ$& 9.0$^\circ$& 0.146&543.3\\
3.0 & 0.67 & 62.7$^\circ$ & -10.8$^\circ$& 0.358&541.1 \\
4.0 & 0.715 & 31.5$^\circ$& -41.5$^\circ$ & 1.05&540.1\\
5.0 & 0.725 & 37.1$^\circ$& -43.7$^\circ$& 1.94&539.6\\
\hline
\end{tabular}}
\caption{Best-fit direction $(l_c, b_c)$ and observer displacement $r_{obs}$ from the center of the inhomogeneity, obtained for the various values of $(r_0,\Omega_{X,\text{in}})$ shown in Fig.~\ref{ontrfig}.}\label{tab1}
\end{table}

In Fig.~\ref{fig5} we show a contour plot of the best-fit value of $r_{obs}$ as a function of the direction in the sky (given by the location of the inhomogeneity center), in galactic coordinates. The largest best-fit value is $r_{obs}=654 {\rm \,Mpc}$ corresponding to the direction $(l_{c,max},b_{c,max})=(336^\circ,-47^\circ)$.

The maximum value of the observer shift $r_{obs}$ consistent with the Union2 data may be obtained by plotting $\chi^2(r_{obs})$ after minimization with respect to $\alpha_c,\delta_c,\mu_0$. This plot is shown in Fig.~\ref{robsmax} for the set of values $(r_0,\Omega_{X,\text{in}})$ shown in Fig.\,\ref{ontrfig}. It shows that even values of $r_{obs}$ as large as $0.7\, r_0$ provide an acceptable fit to the SnIa data i.e. $\chi^2/d.o.f. \simeq 1$ (even though significantly worse than \lcdm, which corresponds to $\chi^2 = 541$). Notice that shifting the observer by $r_{obs}$ results only in a small improvement of the fit ($\Delta\chi^2\lesssim-2$) with respect to the on-center case.

\begin{figure*}[!t]
\centering
\hspace{0pt}\rotatebox{0}{\resizebox{0.8\textwidth}{!}{\includegraphics{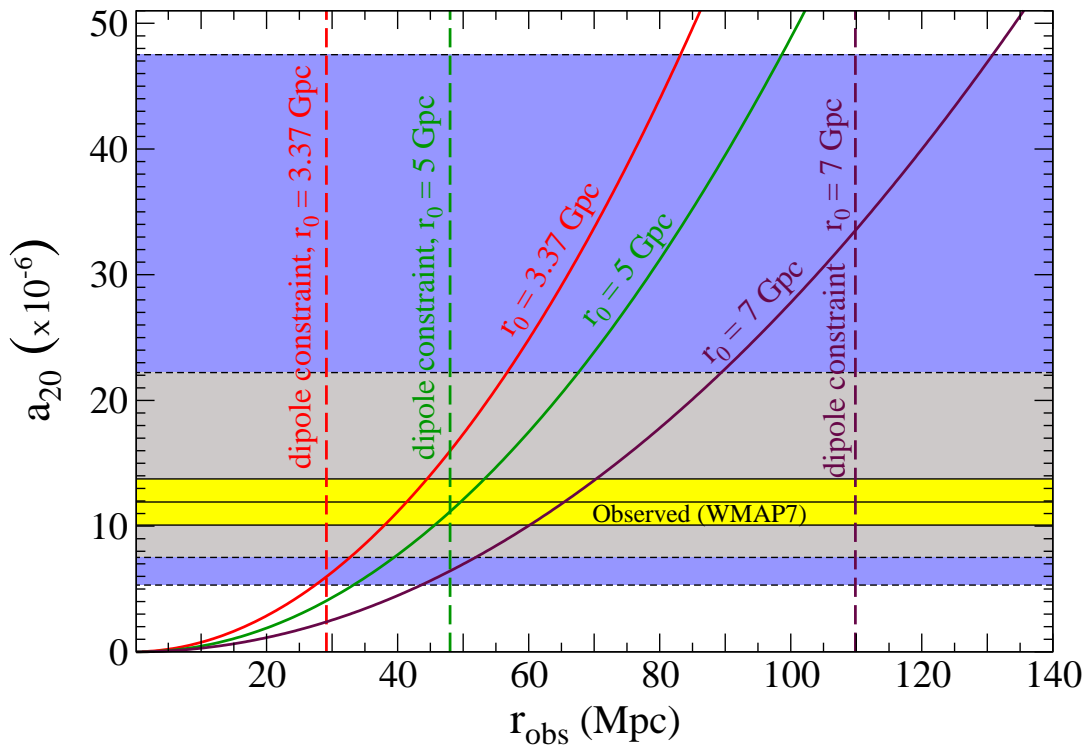}}}
\vspace{0pt}{\caption{a) The quadrupole moment $a_{20}$ as a function of the observer shift $r_{obs}$, for different inhomogeneity sizes $r_0$ and $\Omega_{X,\text{in}}=0.69$. The yellow band represents the value measured by WMAP~\cite{Komatsu:2010fb} including just the observational uncertainty (cf. Eq.~(\ref{quadunc})). The 68\% and 95\% confidence level regions for $a_{20}$ according to the more complete statistical analysis carried out in~\cite{Bennett:2010jb} are also shown (gray and blue shaded regions, respectively). Finally, the vertical dashed lines indicate the maximum value of $r_{obs}$ (for each $r_0$) compatible with the measured dipole, Eq. ($\ref{cobedip2}$).} \label{fig10}}
\end{figure*}

\subsection{Predicted CMB low multipole moments: Constraints from the CMB dipole}

Using Eqs.~(\ref{cmbalm}), (\ref{tempfluct}) it is straightforward to obtain numerically $a_{10}(r_{obs})$, the additional CMB fluctuations dipole moment induced by the shift $r_{obs}$ of the observer. This quantity is shown in Fig.~\ref{fig7} for various values of the inhomogeneity scale $r_0$. In the same plot we show the value of the measured CMB dipole (dashed line)~\cite{Lineweaver:1996qw}
\begin{eqnarray}
\label{measdip}
\left(\frac{\Delta T}{\bar T}\right)_{10}&=&\frac{3.35 \times {10^{-3}}\,{\rm K}}{2.725\,{\rm K}}=(1.230\pm0.013)\times 10^{-3}\nonumber\\
a_{10}&=&\sqrt{\frac{4\pi}{3}}\left(\frac{\Delta T}{\bar T}\right)_{10}\simeq(2.52\pm0.03)\times 10^{-3}\nonumber\\[-3mm]
&&\label{cobedip2}
\end{eqnarray}
Clearly, for $r_0<5{\rm \,Gpc}$ the observer is confined to be in a sphere which is a fine tuned small spatial fraction of the dark energy inhomogeneity ($f(r_0)\equiv(\frac{r_{obs-max}}{r_0})^3\simeq 10^{-6}$) which implies severe fine tuning of the model as in the case of LTB models with matter. This is shown in Fig.~\ref{fig8} where $f(r_0)$ is plotted for values of $r_0$ between 1 and 7 Gpc. As shown in that figure, the fine tuning starts to reduce ($f(r_0)$ increases)  as the size of the inhomogeneity increases beyond $6{\rm \,Gpc}$. We anticipate it to disappear as the size of the inhomogeneity reaches $\sim14$Gpc (the comoving distance to the last scattering surface).
%i.e. for $r_0\simeq 3 H_0^{-1} \simeq 10{\rm \,Gpc}$.

A potentially interesting observation of this class of models is the natural alignment of the low CMB multipoles since the model includes the existence of a preferred axis which connects the observer with the center of the inhomogeneity. The predicted CMB maps corresponding to the dipole quadrupole and octopole are shown in Fig.~\ref{fig9} (b-d) where we have selected the direction of the preferred axis to coincide with the direction of the observed CMB dipole and set $r_{obs}=30$~Mpc. To try to avoid complications related to the numerical precision of the computations (complications that become manifest mainly for low values of $r_{obs}$), we have computed the quadrupole (octopole) in the following way: we first calculated its value (using Eqs.~(\ref{cmbalm}), (\ref{tempfluct})) for 30 values of $r_{obs}$ between 0 and 1.2~Gpc and then performed a quadratic (cubic) fit to those data. The values for $a_{20}$ and $a_{30}$ used in Figs.~\ref{fig9} and \ref{fig10} have been derived from those fits. Notice that the predicted magnitude of the quadrupole is comparable with the observed value. However, the octopole is smaller than the observed value by a factor $\sim 10^{-4}$. Even though the alignment is apparent, the fact that all the $a_{lm}$ coefficients with $m\neq 0$ vanish destroys all the planar features on the plane perpendicular to the preferred axis. Such features could naturally appear if we had considered a {\it lattice} of spherical dark energy inhomogeneities.

In Fig.~\ref{fig10} we show the quadrupole coefficient $a_{20}$  as a function of the observer shift $r_{obs}$. Let us recall that the angular power spectrum is defined through:
\be
\left<a_{lm}a_{l'm'}^*\right>=\delta_{ll'}\delta_{mm'}C_l\,.
\ee
Each $C_l$ can thus be computed by averaging the squares of the corresponding $2l+1$ coefficients $a_{lm}$. However, in our case, due to the axial symmetry (and assuming that the observed anisotropy were due only to our shift from the inhomogeneity center) we have just:
\be
C_l=\frac{\left|a_{l0}\right|^2}{2l+1}
\ee
where, in addition, we should multiply by $\bar{T}=2.725\times10^6\,\mu$K in order to get the $C_l$'s in units of $\mu$K$^2$, which is the usual convention. The yellow shaded region in the plot represents the value measured by WMAP, $C_2=211\pm5\mu{\rm K}^2$~\cite{Komatsu:2010fb}, where the uncertainty only includes the measurement error. This corresponds to:
\begin{equation}
a_{20}=\sqrt{\frac{5 C_2}{(2.725 \times 10^6 \mu K)^2}}\simeq 11.9 \pm 1.8 \label{quadunc}
\end{equation}

We also plot the 68\% (gray) and 95\% (blue) confidence level regions according to the more complete statistical analysis carried out in~\cite{Bennett:2010jb}. 
Notice that the observed value of the quadrupole is reproduced for approximately the same shift of the observer $r_{obs}$ as the one required to produce the observed dipole. This relatively large value of the quadrupole we have derived is in agreement with the results of Ref.~\cite{Tomita:1999rw} but is significantly larger than the value obtained in Ref.~\cite{Alnes:2006pf} for a purely matter void profile.

\section{Discussion and conclusions}

We have considered a generalization of \lcdm by allowing spatial instead of temporal dependence of the dark energy. An advantage of this class of models is that they naturally predict a preferred cosmological direction in agreement with the indications of some cosmological observations~\cite{Copi:2010na,Watkins:2008hf,Hutsemekers:2005iz}. We have considered isocurvature spherically symmetric dark energy inhomogeneities motivated from the possible recent formation of global monopoles with Hubble-scale cores. We have found that such spherical dark energy inhomogeneities with radius less than about $2{\rm \,Gpc}$ are excluded at the $3\sigma$ confidence level by the Union2 SnIa data. Inhomogeneities with larger scales are consistent with the SnIa data and with the CMB dipole provided that the observer remains close to the center of the inhomogeneity, at a distance smaller than a maximum distance $r_{obs-max}$. For the SnIa Union2 data we find $r_{obs-max}\simeq 0.7 \, r_0$ demanding $\chi^2 /d.o.f. \lsim 1$. This is a factor four larger than the corresponding maximum observer shift in a LTB model based on a matter void implying that the fine tuning of the observer location is somewhat reduced in the case of inhomogeneous dark energy. This is partly due to the increase on the size of the observationally allowed inhomogeneities. The corresponding bound obtained from the CMB dipole however is not affected significantly. We find $r_{obs-max}\simeq 110 {\rm \,Mpc}$ even for inhomogeneities as large as $7 {\rm \,Gpc}$. This value of $r_{obs-max}$ increases rapidly as the size of the inhomogeneity approaches the comoving radial distance to the last scattering surface $r_0\simeq 14 {\rm \,Gpc}$. In this limit, the model practically reduces to \lcdm.

Our results have demonstrated that Hubble-scale inhomogeneities of dark energy are consistent with CMB and SnIa cosmological observations provided that the inhomogeneity scale is larger than about $2{\rm \,Gpc}$ and the observer is kept close to the center. Potential future extensions of this project include the following:
\begin{itemize}
\item
Construct a concrete physical model that can give rise to such large-scale dark energy inhomogeneities. Topological monopole configurations discussed in section 2B constitute a possible physical basis for such models.
\item
In the context of the physical model mentioned above, one could try to derive predictions for other cosmological observations. For example, the coupling of electromagnetism or gravity to the inhomogeneous fields of the topological monopole could lead to a natural spatial variation of fundamental constants~\cite{Campanelli:2005gi,Olive:2010vh}  which would be more prominent along the preferred direction that connects the observer to the center of the inhomogeneity. A similar coupling could lead to a spatial coherent rotation of the optical polarization of quasars~\cite{Hutsemekers:2005iz}.
\item
Even though the model does predict an alignment of the low CMB multipoles, this alignment does not affect the multipoles with $m\neq 0$. However, this is no longer true if we consider a superposition of multiple spherical inhomogeneities with scales and distances between their centers of a few ${\rm \,Gpc}$. Such a lattice would keep the existence of a preferred axis and possible introduce concentric cycles in the maps~\cite{Wehus:2010pj,Gurzadyan:2010da} while inducing at the same time alignment between multipoles $a_{lm}$ with $m\neq 0$. The study of the observational predictions of such a lattice would be an interesting extension of this project.
\item
Hubble-scale dark energy inhomogeneities with different symmetries would also be potentially interesting. For example, global vortices would give rise to cylindrically symmetric inhomogeneities while a domain wall would give rise to planar symmetry.
\item
It has been shown that galaxies in the inner region of a spherical inhomogeneity are seen by an off-center observer to have a bulk velocity flow relative to matter outside of the inhomogeneity. The derivation of the predicted profile of such velocity flows in models with inhomogeneous dark energy would also be an interesting extension of the present study.
\end{itemize}
In conclusion, the introduction of spatial variation of dark energy density on Hubble scales constitutes a natural and straightforward generalization of \lcdm which is supported by concrete physical mechanisms. The detailed study of this class of models and its comparison with cosmological observations constitutes an interesting task that has been initiated by the present study.\\

{\bf Numerical Analysis Files:} The \emph{Mathematica} files used to produce the figures are available by e-mail upon request and they will also be available shortly at http://leandros.physics.uoi.gr/ide.\\

\section*{Acknowledgements}
We thank Valerio Marra for useful discussion at the early stages of this work. We have benefited from reading the publicly available code of Ref.~\cite{Marra:2010pg}.


\begin{thebibliography}{99}
%~\cite{Komatsu:2010fb}
\bibitem{Komatsu:2010fb}
  E.~Komatsu {\it et al.}  [WMAP Collaboration],
  %``Seven-Year Wilkinson Microwave Anisotropy Probe (WMAP) Observations:
  %Cosmological Interpretation,''
  Astrophys.\ J.\ Suppl.\  {\bf 192}, 18 (2011)
  [arXiv:1001.4538 [astro-ph.CO]].
  %%CITATION = ARXIV:1001.4538;%%

\bibitem{lcdmrev} P.~J.~E.~Peebles and B.~Ratra,
  %``The cosmological constant and dark energy,''
  Rev.\ Mod.\ Phys.\  {\bf 75}, 559 (2003)
  [arXiv:astro-ph/0207347]; T.~Padmanabhan,
  %``Cosmological constant: The weight of the vacuum,''
  Phys.\ Rept.\  {\bf 380}, 235 (2003)
  [arXiv:hep-th/0212290]; S.~M.~Carroll,
  %``The cosmological constant,''
  Living Rev.\ Rel.\  {\bf 4}, 1 (2001)
  [arXiv:astro-ph/0004075]; V.~Sahni,
  %``The cosmological constant problem and quintessence,''
  Class.\ Quant.\ Grav.\  {\bf 19}, 3435 (2002)
  [arXiv:astro-ph/0202076].

%~\cite{Tegmark:2003ve}
\bibitem{Tegmark:2003ve}
  M.~Tegmark, A.~de Oliveira-Costa and A.~Hamilton,
  %``A high resolution foreground cleaned CMB map from WMAP,''
  Phys.\ Rev.\  D {\bf 68}, 123523 (2003)
  [arXiv:astro-ph/0302496].
  %%CITATION = PHRVA,D68,123523;%%

%~\cite{Land:2005ad}
\bibitem{Land:2005ad}
  K.~Land and J.~Magueijo,
  %``The axis of evil,''
  Phys.\ Rev.\ Lett.\  {\bf 95}, 071301 (2005)
  [arXiv:astro-ph/0502237].
  %%CITATION = PRLTA,95,071301;%%

%~\cite{Copi:2010na}
\bibitem{Copi:2010na}
  C.~J.~Copi, D.~Huterer, D.~J.~Schwarz and G.~D.~Starkman,
  %``Large-angle anomalies in the CMB,''
  Adv.\ Astron.\  {\bf 2010}, 847541 (2010)
  [arXiv:1004.5602 [astro-ph.CO]].
  %%CITATION = 00773,2010,847541;%%

%~\cite{Bennett:2010jb}
\bibitem{Bennett:2010jb}
  C.~L.~Bennett {\it et al.},
  %``Seven-Year Wilkinson Microwave Anisotropy Probe (WMAP) Observations: Are
  %There Cosmic Microwave Background Anomalies?,''
  Astrophys.\ J.\ Suppl.\  {\bf 192}, 17 (2011)
  [arXiv:1001.4758 [astro-ph.CO]].
  %%CITATION = APJSA,192,17;%%

%~\cite{Park:2006dv}
\bibitem{Park:2006dv}
  C.~G.~Park, C.~Park and J.~R.~I.~Gott,
  %``Cleaned Three-Year WMAP CMB Map: Magnitude of the Quadrupole and Alignment
  %of Large Scale Modes,''
  Astrophys.\ J.\  {\bf 660}, 959 (2007)
  [arXiv:astro-ph/0608129].
  %%CITATION = ASJOA,660,959;%%

%~\cite{Abramo:2006gw}
\bibitem{Abramo:2006gw}
  L.~R.~Abramo, A.~Bernui, I.~S.~Ferreira, T.~Villela and C.~A.~Wuensche,
  %``Alignment Tests for low CMB multipoles,''
  Phys.\ Rev.\  D {\bf 74}, 063506 (2006)
  [arXiv:astro-ph/0604346].
  %%CITATION = PHRVA,D74,063506;%%

%~\cite{Gruppuso:2010up}
\bibitem{Gruppuso:2010up}
  A.~Gruppuso and K.~M.~Gorski,
  %``Large scale directional anomalies in the WMAP 5yr ILC map,''
  JCAP {\bf 1003}, 019 (2010)
  [arXiv:1002.3928 [astro-ph.CO]].
  %%CITATION = JCAPA,1003,019;%%

%~\cite{Pontzen:2010yw}
\bibitem{Pontzen:2010yw}
  A.~Pontzen and H.~V.~Peiris,
  %``The cut-sky cosmic microwave background is not anomalous,''
  Phys.\ Rev.\  D {\bf 81}, 103008 (2010)
  [arXiv:1004.2706 [astro-ph.CO]].
  %%CITATION = PHRVA,D81,103008;%%

%~\cite{Rakic:2007ve}
\bibitem{Rakic:2007ve}
  A.~Rakic and D.~J.~Schwarz,
  %``Correlating anomalies of the microwave sky: The Good, the Evil and the
  %Axis,''
  Phys.\ Rev.\  D {\bf 75}, 103002 (2007)
  [arXiv:astro-ph/0703266].
  %%CITATION = PHRVA,D75,103002;%%

%~\cite{Watkins:2008hf}
\bibitem{Watkins:2008hf}
  R.~Watkins, H.~A.~Feldman and M.~J.~Hudson,
  %``Consistently Large Cosmic Flows on Scales of 100 {\rm \,Mpc}/h: a Challenge for the
  %Standard LCDM Cosmology,''
  Mon.\ Not.\ Roy.\ Astron.\ Soc.\  {\bf 392}, 743 (2009)
  [arXiv:0809.4041 [astro-ph]].
  %%CITATION = MNRAA,392,743;%%

%~\cite{Kashlinsky:2008ut}
\bibitem{Kashlinsky:2008ut}
  A.~Kashlinsky, F.~Atrio-Barandela, D.~Kocevski and H.~Ebeling,
  %``A measurement of large-scale peculiar velocities of clusters of galaxies:
  %results and cosmological implications,''
  Astrophys.\ J.\  {\bf 686}, L49 (2009)
  [arXiv:0809.3734 [astro-ph]].
  %%CITATION = ASJOA,686,L49;%%

%~\cite{Lavaux:2008th}
\bibitem{Lavaux:2008th}
  G.~Lavaux, R.~B.~Tully, R.~Mohayaee and S.~Colombi,
  %``Cosmic flow from 2MASS redshift survey: The origin of CMB dipole and
  %implications for LCDM cosmology,''
  Astrophys.\ J.\  {\bf 709}, 483 (2010)
  [arXiv:0810.3658 [astro-ph]].
  %%CITATION = ASJOA,709,483;%%

%~\cite{Hutsemekers:2005iz}
\bibitem{Hutsemekers:2005iz}
  D.~Hutsemekers, R.~Cabanac, H.~Lamy and D.~Sluse,
  %``Mapping extreme-scale alignments of quasar polarization vectors,''
  Astron.\ Astrophys.\  {\bf 441}, 915 (2005)
  [arXiv:astro-ph/0507274];
  D.~Hutsemekers, A.~Payez, R.~Cabanac, H.~Lamy, D.~Sluse, B.~Borguet and J.~R.~Cudell,
  %``Large-Scale Alignments of Quasar Polarization Vectors: Evidence at
  %Cosmological Scales for Very Light Pseudoscalar Particles Mixing with
  %Photons?,''
  arXiv:0809.3088 [astro-ph];
  %%CITATION = ARXIV:0809.3088;%%
  D.~Hutsemekers and H.~Lamy,
  %``Confirmation of the existence of coherent orientations of quasar
  %polarization vectors on cosmological scales,''
  Astron.\ Astrophys.\ {\bf 367}, 381 (2001)
  [arXiv:astro-ph/0012182].
  %%CITATION = ASTRO-PH/0012182;%%

%~\cite{Broadhurst:2004bi}
\bibitem{Broadhurst:2004bi}
  T.~J.~Broadhurst, M.~Takada, K.~Umetsu, X.~Kong, N.~Arimoto, M.~Chiba and T.~Futamase,
  %``The Surprisingly Steep Mass Profile of Abell 1689, from a Lensing Analysis
  %of Subaru Images,''
  Astrophys.\ J.\  {\bf 619}, L143 (2005)
  [arXiv:astro-ph/0412192].
  %%CITATION = ASJOA,619,L143;%%

\bibitem{Umetsu:2007pq}
  K.~Umetsu and T.~Broadhurst,
  %``Combining Lens Distortion and Depletion to Map the Mass Distribution of
  %A1689,''
  Astrophys.\ J.\  {\bf 684}, 177 (2008)
  [arXiv:0712.3441 [astro-ph]].
  %%CITATION = ASJOA,684,177;%%

%~\cite{Klypin:1999uc}
\bibitem{Klypin:1999uc}
  A.~A.~Klypin, A.~V.~Kravtsov, O.~Valenzuela and F.~Prada,
  %``Where are the missing galactic satellites?,''
  Astrophys.\ J.\  {\bf 522}, 82 (1999)
  [arXiv:astro-ph/9901240].
  %%CITATION = ASJOA,522,82;%%

%~\cite{Moore:2001fc}
\bibitem{Moore:2001fc}
  B.~Moore,
  %``The dark matter crisis,''
  AIP Conf.\ Proc.\  {\bf 586}, 73 (2001)
  [arXiv:astro-ph/0103100].
  %%CITATION = ASTRO-PH/0103100;%%

%~\cite{Madau:2008fr}
\bibitem{Madau:2008fr}
  P.~Madau, J.~Diemand and M.~Kuhlen,
  %``Dark matter subhalos and the dwarf satellites of the Milky Way,''
  Astrophys.\ J.\ {\bf 679}, 1260 (2008)
  [arXiv:0802.2265 [astro-ph]].
  %%CITATION = ARXIV:0802.2265;%%

\bibitem{cuspygal}
  G.~Gentile {\it et al.}, %P.~Salucci, U.~Klein, D.~Vergani and P.~Kalberla,
  %``The cored distribution of dark matter in spiral galaxies,''
  Mon.\ Not.\ Roy.\ Astron.\ Soc.\  {\bf 351}, 903 (2004)
  [arXiv:astro-ph/0403154];
  %%CITATION = MNRAA,351,903;%%
  G.~Gentile {\it et al.}, %A.~Burkert, P.~Salucci, U.~Klein and F.~Walter,
  %``The dwarf galaxy DDO 47: testing cusps hiding in triaxial halos,''
  Astrophys.\ J.\  {\bf 634}, L145 (2005)
  [arXiv:astro-ph/0510607];
  %%CITATION = ASJOA,634,L145;%%
  J.~D.~Simon {\it et al.}, %A.~D.~Bolatto, A.~Leroy, L.~Blitz and E.~L.~Gates,
  %``High-Resolution Measurements of the Halos of Four Dark Matter-Dominated
  %Galaxies: Deviations from a Universal Density Profile,''
  Astrophys.\ J.\  {\bf 621}, 757 (2005)
  [arXiv:astro-ph/0412035];
  W.~J.~G.~de Blok,
  %``Halo Mass Profiles and Low Surface Brightness Galaxies Rotation Curves,''
  Astrophys.\ J.\  {\bf 634}, 227 (2005)
  [arXiv:astro-ph/0506753].
  %%CITATION = ASJOA,634,227;%%

%~\cite{Copeland:2006wr}
\bibitem{Copeland:2006wr}
  For a review, see E.~J.~Copeland, M.~Sami and S.~Tsujikawa,
  %``Dynamics of dark energy,''
  Int.~J.~Mod.~Phys. {\bf D15}, 1753 (2006)
  [arXiv:hep-th/0603057].
  %%CITATION = HEP-TH 0603057;%%

\bibitem{stgrav}
  B.~Boisseau, G.~Esposito-Farese, D.~Polarski and A.~A.~Starobinsky,
  %``Reconstruction of a scalar-tensor theory of gravity in an accelerating
  %universe,''
  Phys.\ Rev.\ Lett.\  {\bf 85}, 2236 (2000)
  [arXiv:gr-qc/0001066]; S.~Nesseris and L.~Perivolaropoulos,
  %``The limits of extended quintessence,''
  Phys.\ Rev.\  D {\bf 75}, 023517 (2007)
  [arXiv:astro-ph/0611238]; J.~P.~Uzan,
  %``Cosmological scaling solutions of non-minimally coupled scalar fields,''
  Phys.\ Rev.\  D {\bf 59}, 123510 (1999)
  [arXiv:gr-qc/9903004].

%~\cite{Lue:2005ya}
\bibitem{Lue:2005ya}
  A.~Lue,
  %``The phenomenology of Dvali-Gabadadze-Porrati cosmologies,''
  Phys.\ Rept.\  {\bf 423}, 1 (2006)
  [arXiv:astro-ph/0510068].
  %%CITATION = PRPLC,423,1;%%

\bibitem{frgrav}
  S.~Nojiri and S.~D.~Odintsov,
  %``Dark energy, inflation and dark matter from modified F(R) gravity,''
  in {\it Problems of Modern Theoretical Physics} (TSPU Publishing, Tomsk, 2008), p.266 [arXiv:0807.0685 [hep-th]];   S.~Fay, S.~Nesseris and L.~Perivolaropoulos,
  %``Can $f(R)$ modified gravity theories mimic a $\Lambda$CDM cosmology?,''
  Phys.\ Rev.\  D {\bf 76}, 063504 (2007)
  [arXiv:gr-qc/0703006];   V.~Faraoni,
  %``Solar system experiments do not yet veto modified gravity models,''
  Phys.\ Rev.\  D {\bf 74}, 023529 (2006)
  [arXiv:gr-qc/0607016].
  %%CITATION = PHRVA,D74,023529;%%

%~\cite{Lemaitre:1933qe}
\bibitem{Lemaitre:1933qe}
  G.~Lemaitre,
  Annales Soc.\ Sci.\ Brux.\ Ser.\ I Sci.\ Math.\ Astron.\ Phys.\ A {\bf 53}:51, 1933.
  %%CITATION = ASSBA,A53,51;%%
  For an English translation, see:
  G.~Lemaitre,
  ``The Expanding Universe'',
  Gen.\ Rel.\ Grav.\  {\bf 29}, 641 (1997).
  %%CITATION = GRGVA,29,641;%%

%~\cite{Tolman:1934za}
\bibitem{Tolman:1934za}
  R.~C.~Tolman,
%  ``Effect Of Inhomogeneity On Cosmological Models'',
  Proc.\ Nat.\ Acad.\ Sci.\  {\bf 20}, 169 (1934).
  %%CITATION = PNASA,20,169;%%

%~\cite{Bondi:1947av}
\bibitem{Bondi:1947av}
  H.~Bondi,
%  ``Spherically Symmetrical Models In General Relativity'',
  Mon.\ Not.\ Roy.\ Astron.\ Soc.\  {\bf 107}, 410 (1947).
  %%CITATION = MNRAA,107,410;%%

%~\cite{Krasinski}
\bibitem{Krasinski}
  A.~Krasinski, ``Inhomogeneous Cosmological Models'', Cambridge
  University Press (1997).

%~\cite{Alnes:2005rw}
\bibitem{Alnes:2005rw}
  H.~Alnes, M.~Amarzguioui and O.~Gr\o n,
%  ``An inhomogeneous alternative to dark energy?,''
  Phys.\ Rev.\ D {\bf 73}, 083519 (2006)
 [arXiv:astro-ph/0512006].
  %%CITATION = ASTRO-PH 0512006;%%

%~\cite{Vilenkin:1994pv}
\bibitem{Vilenkin:1994pv}
  A.~Vilenkin,
  %``Topological inflation,''
  Phys.\ Rev.\ Lett.\  {\bf 72}, 3137 (1994)
  [arXiv:hep-th/9402085].
  %%CITATION = PRLTA,72,3137;%%

\bibitem{inprog}J. Grande and L. Perivolaropoulos (in preparation).

%~\cite{Amanullah:2010vv}
\bibitem{Amanullah:2010vv}
  R.~Amanullah {\it et al.},
  %``Spectra and Light Curves of Six Type Ia Supernovae at 0.511 < z < 1.12 and
  %the Union2 Compilation,''
  Astrophys.\ J.\  {\bf 716}, 712 (2010)
  [arXiv:1004.1711 [astro-ph.CO]].
  %%CITATION = ASJOA,716,712;%%

%~\cite{Barrow:1984zz}
%\bibitem{Barrow:1984zz}
  %J.~D.~Barrow and J.~Stein-Schabes,
  %``Inhomogeneous cosmologies with cosmological constant,''
  %Phys.\ Lett.\  A {\bf 103}, 315 (1984).
  %%CITATION = PHLTA,A103,315;%%

%~\cite{Barriola:1989hx}
\bibitem{Barriola:1989hx}
  M.~Barriola and A.~Vilenkin,
  %``Gravitational Field of a Global Monopole,''
  Phys.\ Rev.\ Lett.\  {\bf 63}, 341 (1989).
  %%CITATION = PRLTA,63,341;%%

%~\cite{Kibble:1980mv}
\bibitem{Kibble:1980mv}
  T.~W.~B.~Kibble,
  %``Some Implications Of A Cosmological Phase Transition,''
  Phys.\ Rept.\  {\bf 67}, 183 (1980).
  %%CITATION = PRPLC,67,183;%%

%~\cite{Battye:1998xe}

\bibitem{Battye:1998xe}

  J.~H.~Traschen, N.~Turok and R.~H.~Brandenberger,
  %``MICROWAVE ANISOTROPIES FROM COSMIC STRINGS,''
  Phys.\ Rev.\  D {\bf 34}, 919 (1986);
  %%CITATION = PHRVA,D34,919;%%
  R.~A.~Battye and J.~Weller,
  %``Cosmic structure formation in hybrid inflation models,''
  Phys.\ Rev.\  D {\bf 61}, 043501 (2000)
  [arXiv:astro-ph/9810203].
  %%CITATION = PHRVA,D61,043501;%%
  
%~\cite{Webb:2010hc}
\bibitem{Webb:2010hc}
  J.~K.~Webb, J.~A.~King, M.~T.~Murphy, V.~V.~Flambaum, R.~F.~Carswell and M.~B.~Bainbridge,
  %``Evidence for spatial variation of the fine structure constant,''
  arXiv:1008.3907 [astro-ph.CO].
  %%CITATION = ARXIV:1008.3907;%%

%~\cite{Sakai:1995nh}
\bibitem{Sakai:1995nh}
  N.~Sakai, H.~A.~Shinkai, T.~Tachizawa and K.~i.~Maeda,
  %``Dynamics Of Topological Defects And Inflation,''
  Phys.\ Rev.\  D {\bf 53}, 655 (1996)
  [Erratum-ibid.\  D {\bf 54}, 2981 (1996)]
  %[Phys.\ Rev.\  D {\bf 54}, 2981 (1996)]%
  [arXiv:gr-qc/9506068].
  %%CITATION = PHRVA,D54,2981;%%

%~\cite{Marra:2010pg}
\bibitem{Marra:2010pg}
  V.~Marra and M.~Paakkonen,
  %``Observational constraints on the LLTB model,''
  JCAP {\bf 1012}, 021 (2010)
  [arXiv:1009.4193 [astro-ph.CO]].
  %%CITATION = JCAPA,1012,021;%%

%~\cite{Alnes:2006pf}
\bibitem{Alnes:2006pf}
  H.~Alnes and M.~Amarzguioui,
%  ``CMB anisotropies seen by an off-center observer in a spherically symmetric
%  inhomogeneous universe,''
  Phys.\ Rev.\ D {\bf 74}, 103520 (2006),
  [arXiv:astro-ph/0607334].
  %%CITATION = ASTRO-PH 0607334;%%
  
%~\cite{Enqvist:2006cg}
\bibitem{Enqvist:2006cg}
  K.~Enqvist and T.~Mattsson,
  %``The effect of inhomogeneous expansion on the supernova observations,''
  JCAP {\bf 0702}, 019 (2007)
  [arXiv:astro-ph/0609120].
  %%CITATION = JCAPA,0702,019;%%
  
%~\cite{Tomita:1999rw}
\bibitem{Tomita:1999rw}
  K.~Tomita,
  %``Bulk flows and CMB dipole anisotropy in cosmological void models,''
  Astrophys.\ J.\  {\bf 529}, 26 (2000)
  [arXiv:astro-ph/9905278].
  %%CITATION = ASJOA,529,26;%%

%~\cite{Quercellini:2010zr}
\bibitem{Quercellini:2010zr}
  C.~Quercellini, L.~Amendola, A.~Balbi, P.~Cabella and M.~Quartin,
  %``Real-time Cosmology,''
  arXiv:1011.2646 [astro-ph.CO].
  %%CITATION = ARXIV:1011.2646;%%

%~\cite{Blomqvist:2009ps}
\bibitem{Blomqvist:2009ps}
  M.~Blomqvist and E.~Mortsell,
  %``Supernovae as seen by off-center observers in a local void,''
  JCAP {\bf 1005}, 006 (2010)
  [arXiv:0909.4723 [astro-ph.CO]].
  %%CITATION = JCAPA,1005,006;%%

%~\cite{Alnes:2006uk}
\bibitem{Alnes:2006uk}
  H.~Alnes and M.~Amarzguioui,
  %``The supernova Hubble diagram for off-center observers in a spherically
  %symmetric inhomogeneous universe,''
  Phys.\ Rev.\  D {\bf 75}, 023506 (2007)
  [arXiv:astro-ph/0610331].
  %%CITATION = PHRVA,D75,023506;%%

%~\cite{Blomqvist:2010ky}
\bibitem{Blomqvist:2010ky}
  M.~Blomqvist, J.~Enander and E.~Mortsell,
  %``Constraining dark energy fluctuations with supernova correlations,''
  JCAP {\bf 1010}, 018 (2010)
  [arXiv:1006.4638 [astro-ph.CO]].
  %%CITATION = JCAPA,1010,018;%%

%~\cite{Nesseris:2005ur}
\bibitem{Nesseris:2005ur}
  S.~Nesseris and L.~Perivolaropoulos,
  %``Comparison of the Legacy and Gold SnIa Dataset Constraints on Dark Energy
  %Models,''
  Phys.\ Rev.\  D {\bf 72}, 123519 (2005)
  [arXiv:astro-ph/0511040].
  %%CITATION = PHRVA,D72,123519;%%

%%~\cite{Lineweaver:1996qw}
\bibitem{Lineweaver:1996qw}
  C.~H.~Lineweaver,
  %``The CMB Dipole: The Most Recent Measurement And Some History,''
  in {\it Microwave Background Anisotropies},
  edited by F.~R.~Bouchet {\it et al.}
  (Frontieres, Gif-sur-Yvette, 1997)
  [arXiv:astro-ph/9609034].
  %%CITATION = ASTRO-PH/9609034;%%

%~\cite{Campanelli:2005gi}
\bibitem{Campanelli:2005gi}
  L.~Campanelli, P.~Cea and L.~Tedesco,
  %``Time Variation of the Fine Structure Constant in the Spacetime of a Domain
  %Wall,''
  Mod.\ Phys.\ Lett.\  A {\bf 22}, 1013 (2007)
  [arXiv:astro-ph/0510825].
  %%CITATION = MPLAE,A22,1013;%%

%~\cite{Olive:2010vh}
\bibitem{Olive:2010vh}
  K.~A.~Olive, M.~Peloso and J.~P.~Uzan,
  %``The Wall of Fundamental Constants,''
  Phys.\ Rev.\  D {\bf 83}, 043509 (2011)
  [arXiv:1011.1504 [astro-ph.CO]].
  %%CITATION = PHRVA,D83,043509;%%

%~\cite{Wehus:2010pj}
\bibitem{Wehus:2010pj}
  I.~K.~Wehus and H.~K.~Eriksen,
  %``A search for concentric circles in the 7-year WMAP temperature sky maps,''
  arXiv:1012.1268 [astro-ph.CO].
  %%CITATION = ARXIV:1012.1268;%%

%~\cite{Gurzadyan:2010da}
\bibitem{Gurzadyan:2010da}
  V.~G.~Gurzadyan and R.~S.~Penrose,
  %``Concentric circles in WMAP data may provide evidence of violent
  %pre-Big-Bang activity,''
  arXiv:1011.3706 [astro-ph.CO].
  %%CITATION = ARXIV:1011.3706;%%

\end{thebibliography}
\end{document}